\documentclass[11pt,a4paper]{article}
\usepackage{amsmath}
\usepackage[mathcal]{eucal}
\usepackage{latexsym}
\usepackage{amssymb}
\begin{titlepage}
\title{Finite states in four dimensional quantized gravity.}
\author{Eyo Eyo Ita III}
\medskip
\input amssym.def
\input amssym.tex

\def \in{\indent}

\begin{document}
\maketitle
\bigskip
\centerline{Department of Applied Mathematics and Theoretical Physics} 
\smallskip
\centerline{Centre for Mathematical Sciences, University of Cambridge, Wilberforce Road}
\smallskip
\centerline{Cambridge CB3 0WA, United Kingdom}
\smallskip
\centerline{eei20@cam.ac.uk} 

\bigskip

\begin{abstract}
This is the first in a series of papers outlining an algorithm to explicitly construct finite quantum states of the full theory of gravity in Ashtekar variables.  The algorithm is based upon extending some properties of a special state, the Kodama state for pure gravity with cosmological term, to matter-coupled models.  We then illustrate a presciption for nonperturbatively constructing the generalized Kodama states, in preparation for subsequent works in this series.  We also introduce the concept of the semiclassical-quantum correspondence (SQC).  We express the quantum constraints of the full theory as a system of equations to be solved for the constituents of the `phase' of the wavefunction.  Additionally, we provide a variety of representations of the generalized Kodama states including a generalization of the topological instanton term to include matter fields, for which we present arguments for the field-theoretical analogue of cohomology on infinite dimensional spaces.  We demonstrate that the Dirac, reduced phase space and geometric quantization procedures are all equivalent for these generalized Kodama states as a natural consequence of the SQC.  We relegate the method of the solution to the constraints and other associated ramifications of the generalized Kodama states to separate works.  
\end{abstract}
\end{titlepage}

\section{Introduction}
\par
\medskip
\indent 
The quantization of gravity is currently an unresolved problem in theoretical physics.  A main obstacle to its consistent  quantization lies in the observation that the theory 
of Einstein's gravity, unlike the standard model and quantum chromodynamics, is perturbatively nonrenormalizable in metric variables.  This impasse has led to two main 
alternative approaches to its quantization, namely string theory and loop quantum gravity.  String theory is based upon the idea that Einstein's theory of gravity is the low 
energy limit of a more fundamental theory which should rather be quantized instead, and leads to 26 (or 10, in the case of the superstring) dimensional spacetime.  Loop quantum 
gravity attempts to nonperturbatively quantize gravity in 4-dimensional spacetime in the loop representation and has led to many insights at the kinematic level of the 
gravitational phase space.\par
\indent
One of the ingredients needed for a finite theory of quantum gravity, by the interpretation we adopt in this and in subsequent works, is that given a model coupled to Einstein's relativity in 4-dimensions, one may be able to explicitly construct the physical quantum states devoid of ultraviolet infinities for the full theory as well as for minisuperspace.  A physical quantum state is defined, by this criterion, as a wavefunctional satisfying the quantum version of the constraints of Einstein's general relativity in Ashtekar variables.  To explicitly construct such states one may quantize the theory in the Schr\"odinger representation utilizing Dirac's method for quantizing constrained systems \cite{DIR}.  A good review of the background behind the Ashtekar variables and their resulting simplifications of general relativity can be found in \cite{ROV}, 
\cite{NONPER},\cite{ASH1},\cite{ASH2},\cite{ASH3}.\par
\indent
There is one special state in the full theory of quantum gravity, known as the Kodama state, known to exactly solve the quantum constraints to all orders for a particular operator ordering \cite{KOD}.  The fact that this state as well solves the classical constraints exactly \cite{POSLAMB}, leads to a new conjecture: the principle of the Semiclassical-Quantum correspondence (SQC).  For the pure Kodama state, the SQC amounts to the imposition of a self-duality condition 
constraining the Ashtekar electric and magnetic fields to be proportional to each other by a factor of the cosmological constant $\Lambda$.\par
\indent
The expansion of the Hamiltonian constraint reveals its division into a semiclassical part 
and singular quantum terms.  In the case of pure gravity with $\Lambda$
term, the quantum terms cancel out and the semiclassical part leads directly, via the SQC, to the pure Kodama state.  When matter fields are present in addition to gravity, the SQC 
is broken due to the existence of induced singlar quantum terms.  These singular quantum terms constitute an obstacle to the construction of a finite state of quantum gravity.  Once these singularities are dealt with, then the resulting state once constructed should be finite.  One can then focus on other issues such as normalizability, expectation values, probability currents and reality conditions.  We save treatment of these aspects for future work.\par
\indent
The outline of this paper is as follows.  In section 2 we outline the relevant attributes of the pure Kodama state and introduce the concept of the semiclassical-quantum correspondence, with a view to generalizing these attributes.  In section 3 we generalize the Kodama state by introducing some degrees of freedom (via a CDJ matrix) necessary to incorporate the presence of matter fields quantized with gravity on the same footing.  In section 4 we revisit the kinematic constraints within the context of matter coupling and quantize the constraints subject to the CDJ Ansatz, showing that the SQC remains unbroken.  In section 5 we quantize the Hamiltonian constraint coupled to matter, subject to the CDJ Ansatz.  By requiring that the SQC remain unbroken in spite of the composite nature of the quantum operators involved, we delineate the criteria for finiteness of the generalized Kodama state via a system of nine equations in nine unknowns, shown in section 6.\footnote{The general solution for the CDJ matrix elements in the full theory is beyond the scope of this paper.  The interested reader is directed to \cite{EYOFULL}, where our method of solution is developed in some detail.}  
In section 7 we then delineate the construction of the generalized Kodama state for the full theory from the CDJ matrix elements, explicitly showing its restriction to the final spatial hypersurface $\Sigma_T$ and independence of any field velocities or evolution within the interior of the spacetime manifold $M=\Sigma\times{R}$.  We show the equivalence of the generalized Kodama states $\Psi_{GKod}$ with respect to a variety of representations, including a direct generalization of toplogical field theory to accomodate the presence of the matter fields as well as a via techniques of geometric quantization applied to the cohomology of field theory.\footnote{We relegate the demonstration of the equivalence of the aforementioned representations to the path integral representation to \cite{EYOPATH}, in which we formulate an analogue to the Hartle Hawking noboundary proposal \cite{WAVE1},\cite{WAVE2} in Ashtekar variables.}  It is hoped that this work should consitute a first step in the construction of a finite theory in that we provide a prescription for eliminating untraviolet infinities usually present in the canonical approach to a quantum theory containing composite momentum operators by way of the SQC.  Also, we demonstrate the equivalence of at least three quantization procedures namely: Reduced phase space, Dirac and geometric quantization for the generalized Kodama states.
\par
\indent
We must make a few notes on conventions.  First, our use of the term `generalized Kodama states' is not to be confused with the use in \cite{KODA1} and \cite{KODA2}, where Andrew Randono 
constructs Kodama states for pure gravity using different values of the Immirzi parameter $\gamma$ to label states.  The use of the term in this publication will signify the generalization
from pure gravity to the analogous state when additional fields besides gravity are present 
for $\gamma=\sqrt{-1}$.  Secondly, a quick note on the Ashtekar variables: The basic dynamical variables 
are a left-handed $SU(2)_{-}$ connection, $A^a_i$ and its conjugate momentum, a densitized triad $\widetilde{\sigma}^i_a$ living in a 4-dimensional spacetime manifold
$M=\Sigma\times{R}$.  Our convention for index labeling is that letters from the beginning of the Latin alphabet $a,b,c,...$ signify internal $SU(2)_{-}$ indices and that letters from the middle of the alphabet $i,j,k,...$ signify spatial indices.  Spacetime position in $M$ will be written as $x$, however we will on ocassion highlight the significance of spatial position in $\Sigma$ by use of the boldface $\boldsymbol{x}$, where 
$x=(\boldsymbol{x},t)$.

\section{The pure Kodama state in perspective}
\par
\medskip
\indent

The Einstein-Hilbert action in Ashtekar variables can be written in terms of its 3+1 ADM-type decomposition (\cite{ROV},\cite{NONPER},\cite{ASH1})

\begin{equation}
\label{LAGRAN}
I=\int{dt}\int_{\Sigma}d^3{\boldsymbol{x}}({i \over G}\widetilde{\sigma}^i_{a}\dot{A}^a_i-i\underline{N}H_{grav}-N^{i}(H_i)_{grav}-\theta^{a}G_a),
\end{equation}

\noindent
which is a canonical one-form minus a linear combination of first-class constraints.  The constraints are given by the classical equations of motion for the corresponding Lagrange multipliers, which are nondynamical fields since their time derivatives do not appear in the action.  These are the lapse density
$\underline{N}=N/\sqrt{\hbox{det}h_{ij}}$ (where $h_{ij}$ is the 3-metric), the shift vector 
$N^i$, and the time component of the self-dual Ashtekar connection 
$A^a_0=\theta^a$.  The classical constraints read

\begin{equation}
G_a(x)={{\delta{I}} \over {\delta\theta^{a}(x)}}
=D_{i}\widetilde{\sigma}^i_a(x)=\partial_{i}\widetilde{\sigma}^i_a+f_{abc}A^b_{i}\widetilde{\sigma}^{ci}=0~~\forall{x}\in{M},
\end{equation}

\noindent
which is the $SU(2)_{-}$ Gauss' law constraint with structure constants $f_{abc}$.  Then there is the diffeomorphism constraint

\begin{equation}
(H_i)_{grav}(x)={{\delta{I}} \over  {\delta{N^i(x)}}}=\epsilon_{ijk}\widetilde{\sigma}^j_{a}(x)B^k_{a}(x)=0~~\forall{x}\in{M},
\end{equation}

\noindent
and the Hamiltonian constraint

\begin{equation}
H_{grav}={{\delta{I}} \over {\delta{\underline{N}(x)}}}
= {\epsilon^{abc}}{\epsilon_{ijk}}\widetilde{\sigma}^i_a\widetilde{\sigma}^j_{b}B^k_c
+{1 \over 6}{\epsilon^{abc}}{\epsilon_{ijk}}\widetilde{\sigma}^i_a\widetilde{\sigma}^j_b\widetilde{\sigma}^k_{c}\Lambda=0~~\forall{x}\in{M}.
\end{equation}

These constraints must hold, classically, at all points $x=(\boldsymbol{x},t)$ in the 4-dimensional manifold $M$.  Here, $\boldsymbol{x}$ is the spatial position on a spatial hypersurface $\Sigma_t$ labelled by time $t$.  We are interested in the reduced phase space for this system, which corresponds to the physical degrees of freedom.  The Hamiltonian constraint admits a nontrivial classical solution,
  
\begin{equation}
{{\epsilon^{abc}}{\epsilon_{ijk}}{\widetilde{\sigma}}^i_a{\widetilde{\sigma}}^j_b\Bigl[{\Lambda \over 6}{\widetilde{\sigma}}^k_c}+B^k_c\Bigr]=0
\longrightarrow
{\widetilde{\sigma}}^k_c=-{6 \over \Lambda}B^k_c~~\forall{k,c}
\end{equation}

\noindent
which is the self-duality relation between the Ashtekar electric and magnetic fields, somewhat analogous to the self-duality relation for the electromagnetic field propagating in a vacuum in which $\Lambda^{-1}$ plays the role of $c$, the constant and finite speed of light.  Consistency must be checked with the remaining constraints.

\begin{equation}
G_a=D_{i}\widetilde{\sigma}^i_a=-{6 \over \Lambda}D_{i}B^i_a=0,~~(H_i)_{grav}=\epsilon_{ijk}\widetilde{\sigma}^j_{a}B^k_{a}
=-{6 \over \Lambda}\epsilon_{ijk}B^j_{a}B^k_{a}=0
\end{equation}

\noindent
due to the Bianchi identity and to antisymmetry, respectively.  To evaluate the action on the reduced phase space one substitutes this classical solution back into the starting Lagrangian (\ref{LAGRAN}) yielding

\begin{eqnarray}
\label{EQK}
I=\int{dt}\int_{\Sigma}({i \over G}\widetilde{\sigma}^i_{a}\dot{A}^a_i-i\underline{N}H-N^{i}H_i+\theta^{a}G_a)
\Bigl\vert_{{\widetilde{\sigma}}^k_c=-{6 \over \Lambda}B^k_c}\nonumber\\
=-6i(G\Lambda)^{-1}\int{dt}\int_{\Sigma}B^i_{a}\dot{A}^a_i
\end{eqnarray}

\noindent
on account of the constraints.  Using the identification for the Ashtekar curvature 
$B^i_a=\epsilon^{ijk}\delta_{ae}F^e_{jk}$ where 
$F^e_{jk}=\partial_{j}A^e_{k}-\partial_{k}A^e_j+f^e_{fg}A^f_{j}A^g_{k}$, we can extend this to include the analogous four-dimensional connection by defining 
$F^e_{0i}=\partial_{0}A^e_{i}-\partial_{i}A^e_{0}+f^e_{fg}A^f_{0}A^g_{i}$.  Solving for
$\dot{A}^a_i$ and substituting into (\ref{EQK}) one has

\begin{equation}
\label{PAP1}
I=-6(G\Lambda)^{-1}i\int^T_{t_0}{dt}\int_{\Sigma}d^3{x}\epsilon^{ijk}\delta_{ae}F^e_{jk}(F^a_{0i}+\partial_{i}A^a_{0}-f^a_{fg}A^f_{0}A^g_{i}).
\end{equation}

\noindent
Integrating by parts and dropping boundary terms leads to

\begin{equation}
\label{PAP2}
I=-6(G\Lambda)^{-1}i\int^T_{t_0}{dt}\int_{\Sigma}d^3{x}\epsilon^{ijk}(\delta_{ae}F^e_{jk}F^a_{0i}-A^a_{0}D_{i}B^i_a)
\end{equation}

\noindent
which can be written in covariant notation by defining $\epsilon^{ijk}=\epsilon^{0ijk}$, noting that the second term in brackets in (\ref{PAP2}) vanishes due to the Bianchi 
identity, in the form

\begin{equation}
\label{PAP3}
I=-6(G\Lambda)^{-1}i\int_{M}d^4{x}~\epsilon^{\mu\nu\rho\sigma}F^a_{\mu\nu}F^a_{\rho\sigma}=-6(G\Lambda)^{-1}i\int_{M}\hbox{tr}({F}\wedge{F})
\end{equation}

\noindent
where the trace in (\ref{PAP3}) is taken over left-handed $SU(2)_{-}$ indices.  Let us write the state corresponding to (\ref{PAP3}) in a more recognizable form.  Applying Stokes' theorem 

\begin{equation}
\label{PAP4}
\int_{M}\hbox{tr}({F}\wedge{F})=\int_{\partial_M}L_{CS}[A]=I_{CS}\Bigl\vert_{\partial_{M}},
\end{equation}

\noindent
where $L_{CS}$ is the Chern-Simons action for the left-handed $SU(2)_{-}$ Ashtekar connection living on the boundary $(\Sigma_T,\Sigma_0)\equiv\partial_M$ of $M$, we have

\begin{equation}
I=I_{CS}[A(\Sigma_T)]-I_{CS}[A(\Sigma_0)]
\end{equation}

\noindent
where $I_{CS}$ is the Chern-Simons functional of the $SU(2)_{-}$-valued Ashtekar connection, given by

\begin{equation}
I_{CS}=\int_{\Sigma_T}({A}{dA}+{2 \over 3}{A}\wedge{A}\wedge{A}).
\end{equation}

\noindent
A semiclassical wavefunction can be constructed from this functional by exponentiating $I$, in units of $i/\hbar$, evaluated on the reduced phase space\footnote
{Units of $i/\hbar$ correspond to the Euclidean version of the Kodama state, while units of $1/\hbar$ correspond to the Lorentzian version}.  The exponential of the Chern-Simons functional for quantum gravity is known as the Kodama state, discovered by Hideo Kodama (\cite{KOD},\cite{POSLAMB}), and corresponds to the Hamilton function for the system in Hamilton--Jacobi theory.

\begin{equation}
\label{PAPTOO}
\Psi_{Kod}[A]\propto\hbox{exp}\bigl[-{i \over \hbar}I\bigr]
=\hbox{exp}\bigl[-6(\hbar{G}\Lambda)^{-1}I_{CS}[A]\bigr].
\end{equation}

\indent
Notice how the requirement that the classical constraints be satisfied at all $x$ within $M$ leads to a wavefunctional defined on the 3-dimensional boundary $\Sigma_T=\partial{M}$.  This holographic effect has been demonstrated by Horowitz in \cite{TQFT4}, and is typical of topological field theories.

\subsection{Quantization of the constraints and the semiclassical-quantum correspondence (SQC)}

In order to determine the physical states of quantum gravity the procedure for canonical quantization of constrained systems, developed by Dirac \cite{DIR}, can in some sense be 
used as an alternative to the reduced phase space method introduced below.  In this procedure one promotes the canonically conjugate variables $(A^b_j,\widetilde{\sigma}^i_a)$ to 
quantum operators $(\hat{A}^a_i,\hat{\widetilde{\sigma}}^j_b)$ and Poisson brackets to commutators via the equal-time commutation relations

\begin{eqnarray}
\label{CANCOM}
[A^a_i(\boldsymbol{x},t),{i \over G}\widetilde{\sigma}^j_b(\boldsymbol{y},t)]_{PB}
=\delta^j_{i}\delta^a_b\delta^{(3)}(\boldsymbol{x},\boldsymbol{y})\nonumber\\
\longrightarrow
[\hat{A}^a_i(\boldsymbol{x},t),{i \over G}\hat{\widetilde{\sigma}}^j_b(\boldsymbol{y},t)]
=i\hbar\delta^j_{i}\delta^a_b\delta^{(3)}(\boldsymbol{x},\boldsymbol{y}).
\end{eqnarray}

\noindent
with remaining trivial commutation relations

\begin{eqnarray}
\label{CANCOM100}
[A^a_i(\boldsymbol{x},t),A^b_j(\boldsymbol{y},t)]
=[\widetilde{\sigma}^a_i(\boldsymbol{x},t),\widetilde{\sigma}^b_j(\boldsymbol{y},t)]=0
\end{eqnarray}

\noindent
and defines a Hilbert space for the quantum operators to act on.  To transform the relations (\ref{CANCOM}) into the Schr\"odinger representation one chooses the basis vectors $\bigl\vert{A}^a_i\bigr>$ of the quantum states to be eigenstates of the quantum operator $\hat{A}^a_i(x)$ for a given point $x$.  The state 
$\bigl\vert{A}^a_i\bigr>$ satisfies the orthogonality and completeness relations

\begin{eqnarray}
\label{STATTTES}
\bigl<A^{a_1}_{i_1}(x)\bigl\vert{A}^{\prime{a}_2}_{i_2}(x)\bigr>
=\prod_{\boldsymbol{x}}w[A^{a_1}_{i_1}(x)]^{-1}\delta\bigl(A^{a_1}_{i_1}(x)-{A}^{\prime{a}_2}_{i_2}(x)\bigr)\nonumber\\
\int{D\mu[A]}\bigl\vert{A}\bigr>\bigr<A\bigr\vert
\sim\prod_{\boldsymbol{x},a,i}\int{dA^a_i(x)}w[A(x)]
\bigl\vert{A}^{a_1}_{i_1}(x)\bigr>\bigr<A^{a_1}_{i_1}(x)\bigr\vert=I.
\end{eqnarray}

\noindent
We have not used a gauge invariant, diffeomorphism invariant measure as in loop quantum gravity 
\cite{ROV},\cite{GAMBINI} since our method will be to allow gauge and diffeomorphism invariance of the state to be imposed by the explicit solution of the quantum constraints.  The `weighting' functional $w[A]$ can be chosen judiciously.  A good example for judicious choices of weighting functions in quantum theories was given by Ashtekar and Rovelli in \cite{ASHROV} in Maxwell theory, in which the weighting function for the Bargmann representation was chosen as a measure for the normalization such that certain operators become Hermitian.  We will for the time being leave the weighting function $w[A]$ unspecified, but reserve the freedom to choose it appropriately when the opportunity presents itself.

\par
\indent  
The property of infinite dimensional spaces, which is indigenous to all quantum field theories including quantum gravity, is a direct consequence of consideration of the full theory as opposed to minisuperspace.  Any state $\bigl\vert\Psi\bigr>$ can be expressed in this basis by projecting it onto the complete set of 
states (\ref{STATTTES}) defined on a particular spatial hypersurface $\Sigma_t$ corresponding to time $t$, as presented by Misner in \cite{FEYNQUANT}

\begin{eqnarray}
\bigl\vert\Psi(t)\bigr>
=\int{D\mu[A(t)]}\bigl\vert{A}(t)\bigr>\bigr<{A}(t)\bigr\vert\Psi\bigr>
\end{eqnarray}

\noindent
with  $\bigl<{A}\bigl\vert\Psi\bigr>=\Psi[A]$.  Taking a basis of quantum states in the holomorphic representation of the (three-dimensional) Ashtekar connection 
$\Psi_{Kod}[A]=\bigl<{A}\bigl\vert\Psi\bigr>$, the action of 
$(\hat{A}^a_i,\hat{\widetilde{\sigma}}^j_b)$ are represented respectively by multiplication and functional differentiation

\begin{eqnarray}
\hat{A}^a_i(x)\Psi_{Kod}[A]={A}^a_i(x)\Psi_{Kod}[A]\nonumber\\
\hat{\widetilde{\sigma}}^j_b(x)\Psi_{Kod}[A]
=\hbar{G}{\delta \over {\delta{A^a_i}(x)}}\Psi_{Kod}[A].
\end{eqnarray}

\noindent
According to Dirac, the physical Hilbert space $\Psi_{Phys}$ forms the subset of the full Hilbert space satisfying the quantum version of the constraints, with operator ordering taken into account.  We will attempt to find physical states in the simultaneous kernel of the quantum contraints for an operator ordering with 
the `momenta' to the left of the `coordinate' variables.

\begin{equation}
\hat{G}_a(x)\Psi_{Kod}[A]=\hat{H}_i(x)\Psi_{Kod}[A]=\hat{H}(x)\Psi_{Kod}[A]=0~~\forall{x}.
\end{equation}

\noindent
Using an Ansatz $\Psi_{Kod}[A]=\hbox{exp}\bigl[(G\hbar)^{-1}I[A]\bigr]$, the Gauss' law constraint, which is a statement of the invariance of the quantum state under $SU(2)_{-}$ rotations of the connection, reads

\begin{equation}
{{\delta\Psi_{Kod}[A]} \over {\delta\theta^a(x)}}=\hat{G_a}\Psi
=\hbar{G}D_{i}{{\delta\Psi_{Kod}[A]} \over {\delta{A}^a_i(x)}}
=\Psi_{Kod}[A]{D}_i\Bigl({{\delta{I}} \over {\delta{A}^a_i(x)}}\Bigr)=0.
\end{equation}

\noindent
Note that the quantum condition on the quantum wavefunction $\Psi_{Kod}[A]$ implies an identical condition on its `phase' $I$, which can be viewed as a semiclassical condition.  This constitutes a semiclassical-quantum correspondence for the Gauss' law constraint due to the constraint's being linear in momenta.  The diffeomorphism constraint is a statement of the invariance of the wavefunction under spatial coordinate transformations of its argument $A^a_i(x)$ and reads

\begin{eqnarray}
\label{ANNN}
{{\delta\Psi_{Kod}[A]} \over {\delta{N}^i(x)}}=\hat{H_i}\Psi_{Kod}[A]
=\Bigl[\epsilon_{ijk}\hbar{G}{{\delta} \over {\delta{A}^a_j(x)}}B^j_a(x)\Bigr]\Psi_{Kod}[A]\nonumber\\
=\Bigl[\hbar{G}\epsilon_{ijk}D^{jk}_{aa}+
\hbar{G}\epsilon_{ijk}{{\delta{I}} \over {\delta{A}^a_j(x)}}B^k_a(x)\Bigr]
\Psi_{Kod}[A]=0
\end{eqnarray}

\noindent
where we have used the definition 

\begin{eqnarray}
\label{ANNN1}
D^{ij}_{ab}(x)={{\delta{B}^i_a(x)} \over {\delta{A}^b_j(x)}}={\delta \over {\delta{A}^b_j(x)}}
\epsilon^{ijk}\Bigl[\delta_{ac}\partial_{j}A^c_k(x)+{1 \over 2}f_{abc}A^b_{j}(x)A^c_{k}(x)\Bigr]\nonumber\\
=\epsilon^{ijk}\bigl(\delta_{ab}\partial_{k}\delta^{(3)}(0)+\delta^{(3)}(0)f_{abc}A^c_{k}(x)\bigr).
\end{eqnarray}

\noindent
The application of (\ref{ANNN1}) to (\ref{ANNN}) implies that $D^{jk}_{aa}=0$ due to antisymmetry of the structure constants and that 
$\partial_{i}\delta^{(3)}(0)=0$.\footnote{This latter condition, which differs from conventional functional calculus, is based on the requirement that spatial and functional differentiation must commute \cite{EYOCALCULUS}.}\par 
\indent
The Gauss' law and diffeomorphism constraints, kinematic constraints, do not correspond to physical transformations.  The Hamiltonian constraint is the dynamical constraint, which does encode nontrivial dynamics of the theory in this case since it is at least quadratic in momenta.

\begin{equation}
{{\delta\Psi_{Kod}[A]} \over {\delta\underline{N}(x)}}=\hat{H}\Psi_{Kod}[A]=0.
\end{equation}

\noindent
Expanded out this reads

\begin{equation}
\label{NONTRIVIAL}
{\hbar^2}{G^2}\epsilon^{abc}\epsilon_{ijk}
{\delta \over {\delta{A^a_i}(x)}}{\delta \over {\delta{A^b_j}(x)}}
\Bigl[B^k_c(x)+{{\hbar{G}\Lambda} \over 6}{\delta \over {\delta{A^c_k}(x)}}\Bigr]\hbox{exp}\bigl[(G\hbar)^{-1}I[A]\bigr]=0,
\end{equation}

\noindent
On the one hand, one can see from (\ref{NONTRIVIAL}) that due to the operator ordering chosen there exists a nontrivial solution in which the operator in square brackets annihilates the state, given by

\begin{equation}
\label{PAP6}
{{\delta{I}} \over {\delta{A^c_k}(x)}}=-6(G\Lambda\hbar)^{-1}B^k_c(x)~~\forall{x}\in{M},
\end{equation}

\noindent
from which, if one could `functionally integrate', one could explicitly determine $I$ and construct a wavefunction.  The condition is defined on a particular 3-surface $\Sigma_t$ on which the constraint is evaluated.  Let us contract the left-hand side of (\ref{PAP6}) by the time derivative $\dot{A}^a_i(x)$ and integrate over all 3-space of the manifold $\Sigma$.

\begin{equation}
\label{PAP61}
\int_{\Sigma}d^3{x}{{\delta{I}} \over {\delta{A^a_i}(\boldsymbol{x},t)}}\dot{A}^a_i(\boldsymbol{x},t)
={{dI} \over {dt}}
=-6(G\Lambda\hbar)^{-1}\int_{\Sigma}d^3{x}B^i_a(\boldsymbol{x},t)\dot{A}^a_i(\boldsymbol{x},t)
\end{equation}

\noindent
which is nothing but the definition of the time derivative of a functional of an independent variable $A^a_i$ defined on 3-space in terms of the evolution of the variable.  Recall that for functional variation on the infinite dimensional functional spaces of the type encountered in field theory,

\begin{equation}
\label{PAPTOOTTT}
\delta{I}=\int_{\Sigma}{d^3}x{{\delta{I}} \over {\delta{A^a_i(x)}}}\delta{A^a_i(x)}
\end{equation}

\noindent
It so happens, then, that $dI/dt$ is a total time derivative.  Integrating from $t=t_0$ to $t=T$, one has that the functional $I$ evolves from the initial 
3-surface $\Sigma_0$ to the final 3-surface $\Sigma_{T}$

\begin{equation}
\label{PAP62}
I(T)-I(t_0)
=\int^T_{t_0}{dt}{{dI} \over {dt}}=\int_{M}\hbox{tr}{F}\wedge{F}
=I_{CS}[A(\Sigma_T)]-I_{CS}[(\Sigma_0)],
\end{equation}

\noindent
where we have used the results from (\ref{PAP4}).  One can obtain the same result by 
applying (\ref{PAP6}) as evaluated on the final spatial hypersurface $\Sigma_T$ to (\ref{PAPTOOTTT}), unsupressing the time label to yield

\begin{equation}
\label{PAPTOOT}
\delta{I}_T=\int_{\Sigma}{d^3}x{B^i_a(\boldsymbol{x},T)}\delta{A^a_i(\boldsymbol{x},T)}=\delta{I}_{CS}[A(T)]
\end{equation}

\noindent
and consequently $I_T=I_{CS}[A(T)]$.  One can now write down the solution to the quantum Hamiltonian constraint as

\begin{equation}
\label{PAPTOOT1}
\Psi_{Kod}[A]=\hbox{exp}\bigl[-6(G\Lambda\hbar)^{-1}I_{CS}[A]\bigr].
\end{equation}

\noindent
were in (\ref{PAPTOOT1}) we have suppressed the label $T$ of the spatial hypersurface $\Sigma_T$ forming the boundary $\partial{M}$.  So the quantum 
state (\ref{PAPTOOT1}) and the semiclassically determined states (\ref{PAPTOO}) coincide to all orders with no quantum corrections.  We will define this property, the `Semiclassical-Quantum Correspondence' (SQC).  The usual prescription by which a classical theory gets promoted to its quantum counterpart is a rough rule of thumb which leads to an ambiguity in quantum theories to choose from of order $\hbar$.  The correct quantum theory is fixed by comparison with experiment.  However, we have demonstrated that from the infinite set of possibilities to choose from there is a unique quantum state which coincides with the classical state exactly to all orders, namely the Kodama state.\footnote{We rename this the `pure' Kodama state since it exists for pure gravity with $\Lambda$ term, devoid of 
any matter fields.}\par
\indent
A reasonable question to ask is what physical theories admit a pure Kodama state.  It has been demonstrated \cite{SAN},\cite{NONPER},\cite{TSU1} the construction of such states by
alternate methods for $N=1$ and $N=2$ supergravities in 4-dimensions.  In \cite{TSU2} a canonical analysis was performed for $N=3$ supergravity.  The reason that no such Kodama state seems to have been constructed, to the present author's knowledge, is due to the fact that for $N\geq{3}$ there automatically exist additional lower spin fields which ruin the topological nature of the models that exhibit the SQC.  We hope to ultimately demonstrate, in this series of publications, a new way to extend the SQC to such models.\par
\indent

\subsection{Semiclassical-quantum correspondence for the pure Kodama state}
\par
\medskip
\indent 
The pure Kodama state is the exact solution to the constraints of the full theory with $\Lambda$ term when there are no matter fields present in addition to gravity, given by 
$\Psi_{Kod}=\Psi_{Kod}[A]$.  This can be represented in terms of the self-duality Ansatz

\begin{equation}
\label{QUANT}
\widetilde{\sigma}^i_{a}=-6\Lambda^{-1}\delta_{ae}B^i_{e}.
\end{equation}

\noindent
The pure Kodama state arguably, issues of normalizability aside, can be said to represent a canonical quantization of four dimensional general relativity in the full 
superspace theory, exactly to all orders with no quantum corrections for pure gravity with 
$\Lambda$ term.  Of course $\Psi_{Kod}$ is a special state, which satisfies the self-duality condition and as well the semiclassical-quantum correspondence (SQC).  Equation (\ref{QUANT}) satisfies the SQC since its quantized counterpart yields the same condition to all orders.  The quantized version of (\ref{QUANT}) is given by

\begin{equation}
\label{QUANT1}
\hbar{G}{\delta \over {\delta{A^a_i}(x)}}\Psi_{Kod}[A]=-{6 \over \Lambda}B^i_{a}(x)\Psi_{Kod}[A].
\end{equation}

\section{Generalized Kodama state and the CDJ Ansatz}
\par
\medskip
\indent
We have reviewed the method for the quantization of gravity for pure gravity with cosmological term $\Lambda$ as it relates to the pure Kodama state $\Psi_{Kod}$.  We now outline a method to extend this to the more general case for a nondegenerate magnetic field $B^i_a$, namely gravity coupled to quantized matter fields.  There are a few complications relative to the pure Kodama state which will arise due to the presence of quantized matter fields in the full theory.  The generalized Kodama state contains functional dependence upon the gravitational $A^a_i$ and matter variables $\Psi_{GKod}=\Psi_{GKod}[A,\phi]$, where $\phi=\phi(x)$ represents the matter fields of the model.  First let us write the starting action for gravity coupled to matter, the analogue of (\ref{LAGRAN}), for completeness.

\begin{equation}
\label{LAGRAN1}
I=\int{dt}\int_{\Sigma}d^3{\boldsymbol{x}}\Bigl[({i \over G}\widetilde{\sigma}^i_{a}\dot{A}^a_i+\pi\dot{\phi}-i\underline{N}(H_{grav}+G\Omega)-N^{i}\bigl((H_i)_{grav}+GH_i\bigr)
-\theta^{a}\bigl(G_a+GQ_a\bigr)\Bigr],
\end{equation}

\noindent
where $\Omega$ is the matter contribution to the Hamiltonian constraint, $H_i$ the matter contribution to the diffeomorphism constraint, $Q_a$ the matter contribution to the Gauss' law constraint, and $\pi=\pi(x)$ the conjugate momentum to the matter field $\phi(x)$.  One can then go through the analogous manipulations with respect to (\ref{LAGRAN1}) as in section 2.\par
\indent  
It has been shown by Thiemann in \cite{THIE2} that when matter fields are present in addition to gravity, the constraints can still be solved at the classical level  by use of the CDJ Ansatz

\begin{equation}
\label{QUANT2}
\widetilde{\sigma}^i_{a}(x)=\Psi_{ae}(x)B^i_{e}(x).
\end{equation}

\noindent
In (\ref{QUANT2}) the CDJ matrix $\Psi_{ae}\equiv\Psi_{ae}[A,\phi]$ in general contains nontrivial functional dependence upon the gravitational and the matter variables at each
point $x$.  The CDJ Ansatz (\ref{QUANT2}) can be viewed as a generalization of the self-duality 
condition (\ref{QUANT}) to accomodate the presence of the matter fields, and contains sufficient degrees of freedom to allow solution of the constraints at the classical level \cite{THIE2}.  Our claim in this paper is that since the CDJ Ansatz is linear in gravitational 
momentum $\widetilde{\sigma}^i_a$, it as well satisfies the SQC.  Therefore one should be able to promote 
(\ref{QUANT2}) to its quantized version without breaking this correspondence.  This is the analogue of (\ref{QUANT1}) for the generalized Kodama state $\Psi_{GKod}$

\begin{equation}
\label{QUANT3}
\hbar{G}{\delta \over {\delta{A^a_i}(x)}}\Psi_{GKod}[A,\phi]
=\bigl(\Psi_{ae}(x)B^i_{e}(x)\bigr)\Psi_{GKod}[A,\phi].
\end{equation}

\noindent
In (\ref{QUANT3}) the self duality condition is generalized from the case of a homogeneous isotropic CDJ matrix, and the CDJ matrix in this context now plays the role of a tensor-valued `generalized' inverse cosmological constant.\par
\indent
The equal-time commutation relations can be read off from the phase space structure of (\ref{LAGRAN1}).  In the case of a Klein--Gordon scalar field they read

\begin{equation}
\label{KGCCR}
\bigl[\hat\phi(\boldsymbol{x},t),\hat{\pi}(\boldsymbol{y},t)\bigr]=i\hbar\delta^{(3)}(\boldsymbol{x}-\boldsymbol{y})
\end{equation}

\noindent
along with the `trivial' relations

\begin{equation}
\label{KGCCCR1}
\bigl[\hat\phi(\boldsymbol{x},t),\hat{\phi}(\boldsymbol{y},t)\bigr]=\bigl[\hat\pi(\boldsymbol{x},t),\hat{\pi}(\boldsymbol{y},t)\bigr]=0
\end{equation}

\noindent
amongst the matter variables, and

\begin{equation}
\label{KGCCR1}
\bigl[\hat{\phi}(\boldsymbol{x},t),\hat{A}^i_a(\boldsymbol{y},t)\bigr]=\bigl[\hat\pi(\boldsymbol{x},t),\hat{A}^i_a(\boldsymbol{y},t)\bigr]=0
\end{equation}

\noindent
for `cross' commutation relations, as well as

\begin{equation}
\label{KGCCR2}
\bigl[\hat{\phi}(\boldsymbol{x},t),\hat{\widetilde{\sigma}^i_a}(\boldsymbol{y},t)\bigr]
=\bigl[\hat{\pi}(\boldsymbol{x},t),\hat{\widetilde{\sigma}^i_a}(\boldsymbol{y},t)\bigr]=0
\end{equation}

\noindent
Equation (\ref{KGCCR1}) and (\ref{KGCCR2}) signify the requirement that $\phi(x)$ and $A^a_i(x)$ as well as their conjugate momenta be dynamically independent variables,\footnote{This is the case irrespective of their individual time histories within $M$.} which will be important in the construction of generalized Kodama states.  Hence while these particular commutation relations appear trivial, we will find that they do indeed contain nontrivial physical content.\par
\indent
From this one can as well derive a matter-momentum analogue on the spatial hypersurface $\Sigma_t$ labeled by $t$, using the Schr\"odinger representation, to the CDJ Ansatz.  

\begin{equation}
\label{ANSATZ2}
\hat{\pi}(x)\Psi_{GKod}[A,\phi]
=-i\hbar{\delta \over {\delta\phi(x)}}\Psi_{GKod}[A,\phi]=\pi(x)\Psi_{GKod}[A,\phi],
\end{equation}

\noindent
where $\pi(x)$ is by definition an Ansatz for the action of the operator $\hat\pi$ on the generalized Kodama state 
$\Psi_{GKod}$.\footnote{Not all states are `momentum eigenstates' satisfying equation (\ref{ANSATZ2}).}\par
\indent
We have generalized the basis states in the Schr\"odinger representation to accomodate the presence of the matter fields via the identifications

\begin{eqnarray}
\label{STATTTESONE}
\bigl<A^{a_1}_{i_1}(x),\phi(x)\bigl\vert{A}^{\prime{a}_2}_{i_2}(x),\phi^{\prime}(x)\bigr>
=\prod_{\boldsymbol{x}}W[A^{a_1}_{i_1}(x),\phi(x)]^{-1}\delta\bigl(A^{a_1}_{i_1}(x)-{A}^{\prime{a}_2}_{i_2}(x)\bigr)
\delta\bigl(\phi(x)-\phi^{\prime}(x)\bigr)\nonumber\\
\int{D\mu[A,\phi]}\bigl\vert{A},\phi\bigr>\bigr<A,\phi\bigr\vert
\sim\prod_{\boldsymbol{x},a,i}\int{dA^a_i(x)}d\phi(x)W[A(x),\phi(x)]
\bigl\vert{A}^{a_1}_{i_1}(x),\phi(x)\bigr>\bigr<A^{a_1}_{i_1}(x),\phi(x)\bigr\vert=I.
\end{eqnarray}

\noindent
for some appropriately chosen weighting functional $W=W[A^a_i,\phi]$.\footnote{We demonstrate 
in \cite{EYOPATH} how $W[A,\phi]$ can be chosen such as to establish formal equivalence of the path-integral representation of $\Psi_{GKod}$ to its canonically determined version.}  Any 
state $\bigl\vert\Psi\bigr>$ can now be expressed in this basis by projecting it onto the complete set of states (\ref{STATTTESONE}) defined on a particular spatial hypersurface $\Sigma_t$ corresponding to time $t$

\begin{eqnarray}
\label{STATUS}
\bigl\vert\Psi(t)\bigr>
=\int{D\mu[A(t),\phi(t)]}\bigl\vert{A}(t),\phi(t)\bigr>\bigr<A(t),\phi(t)\bigr\vert\Psi\bigr>
\end{eqnarray}

\noindent
such that $\Psi_{GKod}[A^a_i,\phi]=\bigl<A^a_i,\phi\bigl\vert\Psi\bigr>$.\par
\indent
The CDJ Ansatz applies only when the Ashtekar magnetic field $B^i_a$ is nondegenerate.  A degenerate magnetic field implies a degenerate densitized triad $\widetilde{\sigma}^i_a$, which imples a degenerate 3-metric.  We are not considering in this work states that have support on degenerate metrics.  Note that the existence of matter fields in the theory, which is what distinguishes 
$\Psi_{GKod}$ from $\Psi_{Kod}$ in this paper, implies the nondegeneracy of $B^i_a$ \cite{THIE2}.  This can also be argued from the standpoint that the standard matter Lagrangians require an inverse metric $g^{\mu\nu}$ \cite{DEGENERATE}.\par
\indent

\subsection{Mixed partials condition}

Since the gravitational and matter fields $A^a_{i}(x)$ and $\phi(x)$ respectively are independent dynamical variables, then they must have trivial commutation relations with each other.  So the commutation relations from (\ref{KGCCR2}) read

\begin{equation}
\label{TRIVIAL}
\bigl[\hat{A}^a_i(\boldsymbol{x},t),\hat{\phi}(\boldsymbol{y},t)\bigr]
=\bigl[\hat{\widetilde{\sigma}}^i_a(\boldsymbol{x},t),\hat{\pi}(\boldsymbol{y},t)\bigr]=0~~\forall{\boldsymbol{x},\boldsymbol{y}}
\end{equation}

\noindent
The right hand side of (\ref{TRIVIAL}) in its action on the generalized Kodama state 

\begin{equation}
\label{MIXX}
\bigl[\hat{\widetilde{\sigma}}^i_a(\boldsymbol{x},t),\hat{\pi}(\boldsymbol{y},t)\bigr]\Psi_{GKod}[A^a_i,\phi]=0
\end{equation}

\noindent
leads to a condition known as the mixed partials condition.  Let us proceed into the 
Schr\"odinger representation taking $\boldsymbol{x}=\boldsymbol{y}$

\begin{equation}
\label{MIXX1}
-i\hbar^2{G}\Bigl[{\delta \over {\delta{A^a_i(x)}}}{\delta \over {\delta{\phi(x)}}}
-{\delta \over {\delta{\phi(x)}}}{\delta \over {\delta{A^a_i(x)}}}\Bigr]\Psi_{GKod}=0
\end{equation}

\noindent
Taking the first functional derivatives in (\ref{MIXX1}),

\begin{eqnarray}
\label{MIXX2}
\Bigl[\hbar{G}{\delta \over {\delta{A^a_i(x)}}}(\pi(x))
+i\hbar{\delta \over {\delta{\phi(x)}}}(\Psi_{ae}(x)B^i_{e}(x))\Bigr]\Psi_{GKod}\nonumber\\
=\Bigl[\pi\Psi_{ae}B^i_{e}-\Psi_{ae}B^i_{e}\pi
+\delta^{(3)}(0)\Bigl(\hbar{G}{{\partial\pi} \over {\partial{A}^a_i}}
+i\hbar{B}^i_e{{\partial\Psi_{ae}} \over {\partial\phi}}\Bigr)\Bigr]\Psi_{GKod},
\end{eqnarray}

\noindent
the semiclassical part cancels out and does not lead to anything new.  In order for (\ref{MIXX2}) to be valid, the coefficient of $\delta^{(3)}(0)$ must vanish as well.\footnote{Also, it is assumed that $\Psi_{GKod}$ is an eigenstate of the momentum operator.}  This implies

\begin{eqnarray}
\label{MIXX3}
{{\partial\pi} \over {\partial{A}^a_i}}
=-{i \over G}{B}^i_e{{\partial\Psi_{ae}} \over {\partial\phi}}~~\forall{x}.
\end{eqnarray}

\noindent
Note that (\ref{MIXX3}) is a condition which holds separately at each point $\boldsymbol{x}$ on the 
hypersurface $\Sigma_t$ for each time $t$.  The mixed partials condition will be useful in the elimination of matter momentum in the construction of the generalized Kodama states, and is a key input for the definition of $\Psi_{GKod}$.\par
\indent 
The general solution of (\ref{MIXX3}) can be written, by integration over the functional space of Ashtekar connections $A\in\Gamma$ at fixed position 
$\boldsymbol{x}$, for the specified spatial hypersurface $\Sigma_t$, as\footnote{The brackets around $\int\delta{X}^{ae}$ are meant to signify that the integration constitutes a linear operator, which must be separated from the vector space $\Psi_{ae}$ which it acts on.}

\begin{eqnarray}
\label{MIXX4}
\pi=\pi[A,\phi]=f(\phi)
-{i \over G}{{\partial} \over {\partial\phi}}
\Bigl(\int_{\Gamma}\delta{X}^{ae}\Bigr)\Psi_{ae}[A,\phi],
\end{eqnarray}

\noindent
where $f(\phi)$ is an arbitrary function of the matter field $\phi(x)$ acting as a `constant' of functional integration with respect to the gravitational configuration variables $A^a_i(x)$ for each spatial point $\boldsymbol{x}$, and where the `functional' one-form $\delta{X}^{ae}$ is given by

\begin{eqnarray}
\label{MIXX5}
\delta{X}^{ae}(x)=B^i_e(x)\delta{A}^a_i(x).
\end{eqnarray}

\noindent
There are four things to note concerning (\ref{MIXX4}).  First, it is a linear relation between the matter momentum $\pi(x)$ and the CDJ matrix $\Psi_{ae}(x)$ arising as a consistency condition of the quantization procedure.  Secondly, it has the same functional form for each position $\boldsymbol{x}$ in 
$\Sigma$, and therefore resembles a minisuperspace equation but is in actuality the full theory.  Third, one can consider the allowable semiclassical matter momenta to be labeled by the arbitrary function $f$.  Fourth,  even though $f$ is freely specifiable, it can be judiciously chosen for instance such as to produce a `boundary condition' on the generalized Kodama state such that the proper wavefunction is obtained in the limit when there is no gravity.  The function $f$ then can serve to reproduce the wavefunction which would result from solving the functional Schr\"odinger equation for the Klein--Gordon Hamiltonian in Minkowski 
spacetime.\footnote{Or any appropriate semiclassical limit even which corresponds to observable effects below the Planck scale, if desired.}

\section{Quantization of the generalized kinematic constraints}

\subsection{Diffeomorphism constraint}

The classical diffeomorphism constraint in the presence of matter reads

\begin{equation}
\label{QUANT4}
\epsilon_{ijk}\widetilde{\sigma}^j_{a}B^k_{a}=GH_{i}=G\pi\partial_{i}\phi,
\end{equation}

\noindent
where $H_i$ is the matter contribution.  In (\ref{QUANT4}) we have introduced a factor of $G$ to balance the dimensions in accordance with the time-space part of the Einstein equations $G_{0i}=G{T_{0i}}$.  Substution of the CDJ Ansatz (\ref{QUANT2}) into (\ref{QUANT4}) yields the condition 

\begin{equation}
\label{QUANT5}
\epsilon_{ijk}(\Psi_{ae}B^j_{e})B^k_{a}=GH_{i}.
\end{equation}

\noindent
Using the relation $\epsilon_{ijk}B^j_{e}B^k_{a}=(\hbox{det}B)(B^{-1})^d_{i}\epsilon_{dae}$ and assuming the nondegeneracy of the Ashtekar magnetic field $B^i_a$, we have

\begin{equation}
\label{QUANT5ONE}
\epsilon_{dae}\Psi_{ae}=G{{B^i_{d}H_i} \over {\hbox{det}B}}=G\widetilde{\tau}_{0d},
\end{equation}

\noindent
where $\widetilde{\tau}_{0d}$ is the projection of the time-space component of the matter energy momentum tensor $T_{0i}$ into $SU(2)_{-}$.  The tilde signifies that it is rescaled, or densitized, 
by a factor of $\hbox{det}^{-1}B$, signifying again the nondegeneracy requirement alluded to earlier.  Equation (\ref{QUANT5ONE}) is a statement that the antisymmetric part of the CDJ matrix is uniquely fixed by the matter contribution to the quantum diffeomorphism constraint, as noted in \cite{THIE2}.\par
\indent  
There are two main points of interest regarding (\ref{QUANT5ONE}):  First, the constraint is locally satisfied as a linear relation.  The antisymmetric components of 
$\Psi_{ab}$ here and now depend upon the local matter momentum here and now.  Secondly, a nontrivial right hand side to 
(\ref{QUANT5}) signifies one of the differences between the full superspace and minisuperspace theories since it contains spatial gradients of the matter fields which would otherwise be zero due to spatial homogeneity.  In the minisuperspace theory one would have $H_i=0$ for a 
Klein--Gordon scalar field, corresponding to a symmetric CDJ matrix.\par
\indent
The quantized diffeomorphism constraint reads, taking $H_i$ as the matter contribution, as  

\begin{eqnarray}
\label{ANSATZ4}
\bigl(\epsilon_{ijk}\hat{\widetilde{\sigma}}^j_{a}(x)B^k_a(x)
+G\pi(x)\partial_{i}\phi(x)\bigr)\bigl\vert\Psi_{GKod}\bigr>=\nonumber\\
\Bigl[\hbar{G}\epsilon_{ijk}{\delta \over {\delta{A^b_j(x)}}}B^k_b(x)
-iG\hbar{\delta \over {\delta\phi(x)}}\partial_{i}\phi(x)\Bigr]\bigl\vert\Psi_{GKod}\bigr>
\nonumber\\
=\Bigl(\hbar{G}\delta^{(3)}(0)\epsilon_{ijk}D^{kj}_{bb}
-iG\hbar(\partial_{i}\delta^{(3)}(0))
+\epsilon_{ijk}\Psi_{be}B^j_{e}B^k_{b}
+G\pi\partial_{i}\phi\Bigr)\bigl\vert\Psi_{GKod}\bigr>\nonumber\\
=\Bigl(\epsilon_{ijk}\Psi_{be}B^j_{e}B^k_{b}
+G\pi\partial_{i}\phi\Bigr)\bigl\vert\Psi_{GKod}\bigr>=0.
\end{eqnarray}

\noindent
In (\ref{ANSATZ4}) we have used the quantized CDJ Ansatz as well as the definition 
$D^{ij}_{ab}=\delta{B}^i_{a}/\delta{A}^b_{j}$.  The quantum terms in (\ref{ANSATZ4}) vanish due to antisymmetry of $D^{ij}_{ab}$ and the assumption that functional differentiation and spatial differentiation commute as in (\ref{ANNN1}).\footnote{This assumption differs from conventional calculus.  A brief development of some basic techniques for dealing with field theoretical singularities can be found in \cite{EYOCALCULUS}.}  The requirement that the quantized diffeomorphism constraint be satisfied leads to the condition 

\begin{equation}
\label{DIFF}
\epsilon_{ijk}\Psi_{be}B^j_{e}B^k_{b}+G\pi\partial_{i}\phi=0.
\end{equation}

\noindent
Equation (\ref{DIFF}) is precisely the same condition that would arise from the classical part of the constraint, and therefore satisfies a semiclassical-quantum correspondence.  In fact, this result holds independently of the chosen operator ordering.  It is a property of constraints which are linear in conjugate momenta that the operator ordering for the quantized version is immaterial \cite{TQFT4},\cite{SCHRO6}.

\subsection{Gauss' law constraint}

The classical Gauss' law constraint in the presence of the matter fields reads

\begin{equation}
\label{QUANT6}
D_{i}\widetilde{\sigma}^i_{a}(x)=GQ_a(x).
\end{equation}

\noindent
Again, note that the matter contribution $Q_a$ contains a factor of $G$ relative to the gravitational contribution in (\ref{QUANT6}) in order to balance the dimensions.  Implicit in (\ref{QUANT6}) is a dimensionless constant $\lambda$, which represents the numerical value of the matter $SU(2)_{-}$ charge, in the definition of $Q_a$ 
.  This would be the analogue of the electric charge $e$ in Maxwell theory and is expected to be very small.  Substitution of the CDJ Ansatz into (\ref{QUANT6}) leads to the condition

\begin{equation}
D_{i}(\Psi_{ae}B^i_{e})=\Psi_{ae}D_{i}B^i_{e}+B^i_{e}D_{i}\Psi_{ae}=GQ_a.
\end{equation}

\noindent
Using the Bianchi identity this yields

\begin{equation}
\label{QUANT7}
B^i_{e}D_{i}\Psi_{ae}=GQ_a.
\end{equation}

\noindent
The quantum Gauss' law constraint reads

\begin{equation}
\label{GAUSS1}
\hat{G}_{a}\bigl\vert\Psi_{GKod}\bigr>
=\Bigl[D_{i}\Bigl(\hbar{G}{\delta \over {\delta{A^a_i(x)}}}\Bigr)
+GQ_a(x)\Bigr]\bigl\vert\Psi_{GKod}\bigr>=0.
\end{equation}

\noindent
where $Q_a$ is the $SU(2)_{-}$ charge for a general matter field, given by

\begin{eqnarray}
Q_{a}(x)=\lambda\pi_{\alpha}(x)(T_{a})^{\alpha}_{\beta}\phi^{\beta}(x).
\end{eqnarray}

\noindent
Incorporating the matter contribution and applying the CDJ Ansatz, the matter constraint reads

\begin{eqnarray}
\label{GOSS}
\Bigl[D_{i}(\Psi_{ae}(x)B^i_{e}(x))
-i\hbar\lambda{G}{{\delta\phi^{\beta}(x)} \over {\delta\phi^{\alpha}(x)}}(T_a)^{\alpha}_{\beta}
-i\hbar\lambda{G}\phi^{\beta}
(T_a)^{\alpha}_{\beta}{\delta \over {\delta\phi^{\alpha}(x)}}\Bigr]\Psi_{GKod}\nonumber\\
=\Bigl(D_{i}(\Psi_{ae}B^i_{e})
-i\hbar\lambda{G}\delta^{(3)}(0)\delta^{\beta}_{\alpha}(T_a)^{\alpha}_{\beta}
+G\lambda\pi_{\alpha}(x)(T_a)^{\alpha}_{\beta}\phi^{\beta}(x)\Bigr)\Psi_{GKod}\nonumber\\
=\Bigl(D_{i}(\Psi_{ae}B^i_{e})+GQ_a\Bigr)\Psi_{GKod}
\end{eqnarray}

\noindent
where we have used in the last line of (\ref{GOSS}) that the $SU(2)_{-}$ charge generators $(T_a)^{\alpha}_{\beta}$ are traceless in order to get rid of the singular $\delta^{(3)}(0)$ term in the second line of (\ref{GOSS}).  So we see from (\ref{GOSS}) that the quantum condition precisely implies the semiclassical condition.  The Gauss' law constraint satisfies the SQC due to being linear in conjugate momenta.\par
\indent
For a Klein-Gordon scalar field $\phi$ the source term $Q_a$ is zero since the field is a Lorentz scalar and therefore does not transform under $SU(2)$.  The Gauss' law constraint for the general case of matter coupled to gravity is given, by the coefficient of $\Psi_{GKod}$ in the last line 
of (\ref{GOSS}), by

\begin{equation}
\label{GOSS1}
D_{i}(\Psi_{ae}B^i_{e})+GQ_a=0
\end{equation}

\noindent
Let us focus first on the first term on the left hand side of (\ref{GOSS1}).  This reads

\begin{eqnarray}
\label{GOSS2}
D_{i}(\Psi_{ae}B^i_{e})=(D_{i}B^i_e)\Psi_{ae}+B^i_{e}D_{i}\Psi_{ae}=B^i_{e}D_{i}\Psi_{ae},
\end{eqnarray}

\noindent
where we have used the Bianchi identity.  The last term of (\ref{GOSS2}) forms the covariant derivative of 
a $SU(2)_{-}\otimes{SU}(2)_{-}$-valued tensor $\Psi_{ae}$.  Expanding this, applying the tensor representation of the covariant derivative, we have

\begin{eqnarray}
\label{GOSS3}
B^i_{e}D_{i}\Psi_{ae}
=B^i_{e}\bigl(\partial_{i}\Psi_{ae}+f_{afg}A^f_{i}\Psi_{ge}+f_{efg}A^f_{i}\Psi_{ag}\bigr)\nonumber\\
=B^i_{e}\partial_{i}\Psi_{ae}+B^i_{e}A^f_{i}\bigl(f_{afg}\Psi_{ge}+f_{efg}\Psi_{ag}\bigr).
\end{eqnarray}

\noindent
Note that for the case of a homogeneous and isotropic CDJ matrix 
$\Psi_{ab}=6\Lambda^{-1}\delta_{ab}$ (e.g. the pure Kodama state), 
(\ref{GOSS3}) would be zero.  Hence factoring out $\Lambda$ which is a numerical constant, we would have

\begin{eqnarray}
\label{GOSS4}
B^i_{e}\partial_{i}\Psi_{ae}+B^i_{e}A^f_{i}\bigl(f_{afg}\Psi_{ge}+f_{efg}\Psi_{ag}\bigr)\nonumber\\
=B^i_{e}\partial_{i}\delta_{ae}+B^i_{e}A^f_{i}\bigl(f_{afg}\delta_{ge}+f_{efg}\delta_{ag}\bigr)\nonumber\\
=0+B^i_{e}A^f_{i}\bigl(f_{afe}+f_{efa}\bigr)=0
\end{eqnarray}

\noindent
due to antisymmetry of the structure constants.\footnote{We provide an explicit general solution to the Gauss' law constraint, in the full theory coupled to matter fields, in \cite{EYOGAUSS}.}

\subsection{Recapitulation of the kinematic constraints}

There are a few items of note regarding the kinematic constraints.\par
\noindent
(i) Both sets of constraints are linear in conjugate momenta and their solution depends linearly upon the matter source, namely the Noether charges corresponding to the respective kinematic symmetries.  The kinematic constraints by definition satisfy the SQC since the operator ordering for the quantized version is immaterial.\par
\noindent
(ii) As a corollary to (i), the processes of Dirac quantization and phase space reduction are the same for the kinematic constraints \cite{TQFT4}.\par
\noindent
(iii) The solutions to the six kinematic constraints eliminate, modulo boundary conditions on the Gauss' law constraints, six out of nine degrees of freedom of the CDJ matrix $\Psi_{ab}$, leaving three degrees of freedom remaining for the Hamiltonian constraint.\par
\noindent
(iv) The diffeomorphism constraint determines the antisymmetric part of the CDJ matrix and depends locally upon the spatial gradients of the matter field, which distinguishes one aspect of the full theory from minisuperspace.  The Gauss' law constraint in general should determine three alternate CDJ matrix elements, reducing the CDJ matrix $\Psi_{ae}$ by an additional three degrees of freedom.  The matter contribution to this constraint is of exactly the same form in the minisuperspace and in the full theory, but does distinguish $SU(2)_{-}$ scalars from fields transforming nontrivially under $SU(2)_{-}$.  The diffeomorphism constraint makes this distinction as well, via the difference between a spatial gradient $\partial_i$ and a 
$SU(2)_{-}$ covariant derivative $D_i=\partial_{i}+A_{i}$.\par
\noindent
(v) Lastly, the quantized versions of the kinematic constraints do not produce any information not already contained in their classical counterparts due to the SQC.  The form of these constraints is model-independent, since they are expressed entirely as a representation of the kinematic gauge algebra, namely a semidirect product of $SU(2)_{-}$ gauge transformations with diffeomorphisms.\footnote{Some interesting relationships between gauge transformations and diffeomorphisms can be found by the interested reader in \cite{EYOKINEMATIC}.}

\section{Quantization of the Hamiltonian constraint}
\par
\medskip
\noindent 

The classical Hamiltonian constraint is given by

\begin{equation}
\label{QUIT}
{\Lambda \over 6}\epsilon_{ijk}\epsilon^{abc}\widetilde{\sigma}^i_{a}\widetilde{\sigma}^j_{b}\widetilde{\sigma}^k_{c}
+\epsilon_{ijk}\epsilon^{abc}\widetilde{\sigma}^i_{a}\widetilde{\sigma}^j_{b}B^k_{c}+G\Omega=0,
\end{equation}

\noindent
where $\Omega$ is the matter contribution.  Substitution of the classical CDJ Ansatz results in one condition relating the invariants of the CDJ matrix, namely

\begin{equation}
\label{QUIT1}
\hbox{det}B\bigl(Var\Psi+\Lambda\hbox{det}\Psi\bigr)+G\Omega=0.
\end{equation}

\noindent
where we have defined $Var\Psi$, the variance of the CDJ matrix, as

\begin{eqnarray}
Var\Psi=(\hbox{tr}\Psi)^2-\hbox{tr}\Psi^2.
\end{eqnarray}

\noindent
The matter contribution in general contains dependence upon the CDJ matrix elements in addition to the matter fields $\Omega=\Omega[\phi,\pi,\Psi_{ab}]$, due to the gravity-matter coupling 
$g^{\mu\nu}T_{\mu\nu}$ stemming from the Einstein equations $G_{\mu\nu}\propto{T}_{\mu\nu}$.  Unlike for the kinematic constraints, the form of the matter contribution to the Hamiltonian constraint is model-specific.\par
\indent
Solution of (\ref{QUIT1}) would eliminate one degree of freedom in the CDJ matrix.  Combined with the six kinematic constraint solutions this leaves, modulo boundary conditions due to the Gauss' law constraint, two degrees of freedom remaining in the CDJ matrix $\Psi_{ab}$.  This results in a two-parameter ambiguity in the solution at the classical level.\par
\indent
Since we are interested in generalized quantum Kodama states, we must solve the quantized version of the Hamiltonian constraint.  The quantized version 
of (\ref{QUIT}) is given, with respect to the spatial hypersurface $\Sigma_T$ labelled by $T$, by

\begin{eqnarray}
\label{QUIT2}
\hat{H}\Psi_{GKod}[A^a_i,\phi^{\alpha}]=
\biggl[{\Lambda \over 6}\hbar^3{G^3}{\epsilon^{abc}}{\epsilon_{ijk}}{\delta \over {\delta{A^a_i}(x)}}{\delta \over {\delta{A^b_j}(x)}}{\delta \over {\delta{A^c_k}(x)}}\nonumber\\
+\hbar^2{G^2}{\epsilon^{abc}}{\epsilon_{ijk}}B^i_{a}{\delta \over {\delta{A^b_j}(x)}}{\delta \over {\delta{A^c_k}(x)}}
+\hat{\Omega}\bigl[\phi^{\alpha},\delta/\delta\phi^{\alpha},\delta/\delta{A^a_i}\bigr]\biggr]\Psi_{GKod}[A,\phi]=0~\forall{x}.
\end{eqnarray}

\noindent
A question arises as to whether the SQC is broken due to (\ref{QUIT2}) being cubic in conjugate momenta unlike for the kinematic constraints.  Substitution of the quantized CDJ Ansatz (\ref{QUANT3}) into (\ref{QUIT2}) leads to a condition of the form (suppressing the implicit $T$ dependence)

\begin{equation}
\label{SING}
\hat{H}({x})\bigl\vert\Psi_{GKod}\bigr>=\bigl(q_0(x)+\hbar{G}\delta^{(3)}(\boldsymbol{x})q_{1}(x)
+(\hbar{G}\delta^{(3)}(\boldsymbol{x}))^{2}q_{2}(x)\bigr)\bigl\vert\Psi_{GKod}\bigr>=0
\end{equation}

\noindent
for all $\boldsymbol{x}$ in the spatial hypersurface $\Sigma_T$, whereupon the nonlinear action of the Hamiltonian constraint upon the state leads to the presence of singular quantum gravitational terms.\par
\indent  
In a usual field-theoretical treatment, such terms would be regularized by some prescription in order to yield a finite result.  However, there is no guarantee that the result obtained by solving the regularized constraint is independent of the regularization prescription \cite{GAMBINI}.  Therefore we shall dispense with any regularization procedures altogether in the canonical part of our quantum treatment of gravity.  Instead, we will show that, due to the choice of the generalized Kodama state, singular terms in the constraint equations are cancelled out.\par
\indent

\subsection{Cosmological contribution to the expansion of the quantized Hamiltonian constraint}

There are a total of three contributions to the quantized Hamiltonian constraint namely the cosmological term, the curvature and the matter terms.  Starting with the cosmological term $H_{\Lambda}$, suppressing the implicit $x$ dependence,

\begin{eqnarray}
\label{HAMMI}
\hat{H}_{\Lambda}\bigl\vert\Psi_{GKod}\bigr>=
{\Lambda \over 6}(\hbar{G})^3\epsilon_{ijk}\epsilon^{abc}{\delta \over {\delta{A}^a_i}}{\delta \over {\delta{A}^b_j}}{\delta \over {\delta{A}^c_k}}\bigl\vert\Psi_{GKod}\bigr>\nonumber\\
={\Lambda \over 6}(\hbar{G})^2\epsilon_{ijk}\epsilon^{abc}{\delta \over {\delta{A}^a_i}}{\delta \over {\delta{A}^b_j}}(\Psi_{ce}B^k_{e})\bigl\vert\Psi_{GKod}\bigr>
\end{eqnarray}

\noindent
In (\ref{HAMMI}) we have made the CDJ Ansatz $\widetilde{\sigma}^k_{c}=\Psi_{ce}B^k_{e}$.  Note
from (\ref{HAMMI}) that the CDJ matrix $\Psi_{ce}$ now plays the role of an inverse `generalized' cosmological constant.  Although it is in general a
field-dependent $SU(2)\otimes{SU}(2)$-valued tensor, it is analogous to $\Lambda^{-1}$ for the pure Kodama state and has mass dimensions $[\Psi_{ce}]=-2$.  Since the Ashtekar magnetic field $B^k_{e}$ has mass dimensions $[B^k_{e}]=2$ it provides a check on dimensional consistency to note that the densitized triad 
$\widetilde{\sigma}^k_{c}$ must be dimensionless $([\widetilde{\sigma}^k_{c}]=0)$.  Continuing along from (\ref{HAMMI}),

\begin{eqnarray}
\label{LAS}
{\Lambda \over 6}(\hbar{G})^2\epsilon_{ijk}\epsilon^{abc}{\delta \over {\delta{A}^a_i}}{\delta \over {\delta{A}^b_j}}(\Psi_{ce}B^k_{e})\bigl\vert\Psi_{GKod}\bigr>\nonumber\\
={\Lambda \over 6}\epsilon_{ijk}\epsilon^{abc}\hbar{G}{\delta \over {\delta{A}^a_i}}
\Bigl[\hbar{G}\delta^{(3)}(\boldsymbol{x}){\partial \over {\partial{A^b_j}}}(\Psi_{ce}B^k_{e})+\Psi_{ce}\Psi_{bf}B^k_{e}B^j_{f}\Bigr]\bigl\vert\Psi_{GKod}\bigr>\nonumber\\
={\Lambda \over 6}\epsilon_{ijk}\epsilon^{abc}\biggl[(\hbar{G}\delta^{(3)}
(\boldsymbol{x}))^{2}{\partial \over {\partial{A^a_i}}}{\partial \over {\partial{A^b_j}}}
(\Psi_{ce}B^k_{e})
+\hbar{G}\delta^{(3)}(\boldsymbol{x}){\partial \over {\partial{A^a_i}}}
(\Psi_{ce}\Psi_{bf}B^k_{e}B^j_{f})\nonumber\\
+\hbar{G}\delta^{(3)}(\boldsymbol{x})(\Psi_{af}B^i_{f})
{\partial \over {\partial{A^b_j}}}(\Psi_{ce}B^k_{e})+\Psi_{ce}\Psi_{bf}\Psi_{ag}B^k_{e}B^j_{f}B^i_{g}\biggr]\bigl\vert\Psi_{GKod}\bigr>.
\end{eqnarray}

\noindent
The observation that the second and third terms in the last two lines of (\ref{LAS}) are proportional to each other enables a simplification of the `eigenvalue' of the cosmological term.  This can be seen by expanding out the coefficient of the quantum gravitational singularity and reshuffling indices.

\begin{eqnarray}
\label{HAMMI1}
{\Lambda \over 6}\epsilon_{ijk}\epsilon^{abc}{\partial \over {\partial{A^a_i}}}
(\Psi_{ce}\Psi_{bf}B^k_{e}B^j_{f})
+{\Lambda \over 6}\epsilon_{ijk}\epsilon^{abc}(\Psi_{af}B^i_f){\partial \over {\partial{A^b_j}}}
(\Psi_{ce}B^k_{e})\nonumber\\
={\Lambda \over 6}\epsilon_{ijk}\epsilon^{abc}(\Psi_{ce}B^k_{e})
{\partial \over {\partial{A^a_i}}}(\Psi_{bf}B^j_{f})
+{\Lambda \over 6}\epsilon_{ijk}\epsilon^{abc}(\Psi_{bf}B^j_{f})
{\partial \over {\partial{A^a_i}}}(\Psi_{ce}B^k_{e})\nonumber\\
+{\Lambda \over 6}\epsilon_{ijk}\epsilon^{abc}(\Psi_{af}B^i_{f})
{\partial \over {\partial{A^b_j}}}(\Psi_{ce}B^k_{e})
\end{eqnarray}

\noindent
in (\ref{HAMMI1}) we have used the Liebnitz rule.  To show all three terms on the right hand side of (\ref{HAMMI1}) are equal, relabel $b\leftrightarrow{c}$, $f\leftrightarrow{e}$ and 
$j\leftrightarrow{k}$ on the second term and $a\leftrightarrow{c}$, $f\leftrightarrow{e}$ and 
$i\leftrightarrow{k}$ on the third term.  This leads to

\begin{eqnarray}
\label{HAMMI2}
{\Lambda \over 6}\epsilon_{ijk}\epsilon^{abc}(\Psi_{ce}B^k_{e}){\partial \over {\partial{A^a_i}}}(\Psi_{bf}B^j_{f})
+{\Lambda \over 6}\epsilon_{ijk}\epsilon^{abc}(\Psi_{bf}B^j_{f}){\partial \over {\partial{A^a_i}}}(\Psi_{ce}B^k_{e})\nonumber\\
+{\Lambda \over 6}\epsilon_{ijk}\epsilon^{abc}(\Psi_{af}B^i_{f}){\partial \over {\partial{A^b_j}}}(\Psi_{ce}B^k_{e})
=
{\Lambda \over 6}\epsilon_{ijk}\epsilon^{abc}(\Psi_{ce}B^k_{e}){\partial \over {\partial{A^a_i}}}(\Psi_{bf}B^j_{f})\nonumber\\
+{\Lambda \over 6}\epsilon_{ijk}\epsilon^{acb}(\Psi_{ce}B^k_{e}){\partial \over {\partial{A^a_i}}}(\Psi_{bf}B^j_{f})
+{\Lambda \over 6}\epsilon_{kji}\epsilon^{cba}(\Psi_{ce}B^k_{e}){\partial \over {\partial{A^b_j}}}(\Psi_{af}B^i_{f})
\end{eqnarray}

\noindent
Relabeling $j\leftrightarrow{i}$ and $a\leftrightarrow{b}$ on the last term on the right hand side of (\ref{HAMMI2}), we obtain a final result for the second and the third terms on the right hand side on the bottom two lines of (\ref{LAS}) which constitute the first-derivative terms of the constraint, of

\begin{eqnarray}
\label{HAMMI3}
{\Lambda \over 6}\epsilon_{ijk}\epsilon^{abc}(\Psi_{ce}B^k_{e}){\partial \over {\partial{A^a_i}}}(\Psi_{bf}B^j_{f})
+{\Lambda \over 6}\epsilon_{ijk}\epsilon^{acb}(\Psi_{ce}B^k_{e}){\partial \over {\partial{A^a_i}}}(\Psi_{bf}B^j_{f})\nonumber\\
+{\Lambda \over 6}\epsilon_{kji}\epsilon^{cba}(\Psi_{ce}B^k_{e}){\partial \over {\partial{A^b_j}}}(\Psi_{af}B^i_{f})
={\Lambda \over 6}\epsilon_{ijk}\epsilon^{abc}(\Psi_{ce}B^k_{e}){\partial \over {\partial{A^a_i}}}(\Psi_{bf}B^j_{f})\nonumber\\
+{\Lambda \over 6}\epsilon_{ijk}\epsilon^{acb}(\Psi_{ce}B^k_{e}){\partial \over {\partial{A^a_i}}}(\Psi_{bf}B^j_{f})
+{\Lambda \over 6}\epsilon_{kij}\epsilon^{cab}(\Psi_{ce}B^k_{e}){\partial \over {\partial{A^b_j}}}(\Psi_{bf}B^j_{f})\nonumber\\
={3 \over 2}\Bigl({\Lambda \over 6}\Bigr)\epsilon_{ijk}\epsilon^{abc}{\partial \over {\partial{A^a_i}}}(\Psi_{ce}\Psi_{bf}B^k_{e}B^j_{f})
={\Lambda \over 4}{\partial \over {\partial{A^a_i}}}(\epsilon_{ijk}\epsilon^{abc}
\Psi_{ce}\Psi_{bf}B^k_{e}B^j_{f}).
\end{eqnarray}

\noindent
The semiclassical part of (\ref{LAS}) is given by

\begin{equation}
\label{SEM}
{\Lambda \over 6}\epsilon_{ijk}\epsilon^{abc}B^k_{e}B^j_{f}B^i_{g}\Psi_{ce}\Psi_{bf}\Psi_{ag}
={\Lambda \over 6}(\hbox{det}B)\hbox{det}\Psi\epsilon_{gfe}\epsilon_{gfe}\nonumber\\
=\Lambda(\hbox{det}B)\hbox{det}\Psi
\end{equation}

\noindent
The factor of $6$ due to the definition of the determinant has cancelled out.  So the total contribution due to the eigenvalue of the cosmological term is given by

\begin{eqnarray}
\label{CURVHAM}
\hat{H}_{\Lambda}\bigl\vert\Psi_{GKod}\bigr>=
{\Lambda \over 6}(\hbar{G})^3\epsilon_{ijk}\epsilon^{abc}{\delta \over {\delta{A}^a_i}}
{\delta \over {\delta{A}^b_j}}{\delta \over {\delta{A}^c_k}}\bigl\vert\Psi_{GKod}\bigr>\nonumber\\
=\biggl[\Lambda(\hbox{det}B)\hbox{det}\Psi
+\hbar{G}\delta^{(3)}(\boldsymbol{x})\Bigl({\Lambda \over 4}{\partial \over {\partial{A^a_i}}}(\epsilon_{ijk}\epsilon^{abc}\Psi_{ce}\Psi_{bf}B^k_{e}B^j_{f})\Bigr)\nonumber\\
+(\hbar{G}\delta^{(3)}(\boldsymbol{x}))^{2}
\Bigl({\Lambda \over 6}\epsilon_{ijk}\epsilon^{abc}
{\partial \over {\partial{A^a_i}}}{\partial \over {\partial{A^b_j}}}
(\Psi_{ce}B^k_{e})\Bigr)\biggr]\bigl\vert\Psi_{GKod}\bigr>
\end{eqnarray}

\noindent
We shall now move on to the curvature contribution,

\subsection{Curvature contribution to the expansion of the quantized Hamiltonian constraint}
\par
\medskip
\noindent 
The curvature contribution $\hat{H}_{curv}$ to the quantized Hamiltonian constraint is given by

\begin{eqnarray}
\label{CURV}
\hat{H}_{curv}\bigl\vert\Psi_{GKod}\bigr>=
(\hbar{G})^2\epsilon_{ijk}\epsilon^{abc}{\delta \over {\delta{A}^a_i}}
{\delta \over {\delta{A}^b_j}}B^k_{c}\bigl\vert\Psi_{GKod}\bigr>\nonumber\\
=\epsilon_{ijk}\epsilon^{abc}\hbar{G}{\delta \over {\delta{A}^a_i}}
\Bigl[\hbar{G}\delta^{(3)}(\boldsymbol{x})D^{kj}_{cb}+B^k_{c}(\Psi_{be}B^j_{e})\Bigr]\bigl\vert\Psi_{GKod}\bigr>
\end{eqnarray}~~~~~~~~~~~~~

\noindent
Note that we have maintained an operator ordering with momenta to the left of the coordinates in analogy to that determining the pure Kodama state.  It has been demonstrated
that for this ordering, the quantum algebra of constraints formally closes \cite{ASH2}, 
\cite{ASH3}.  Continuing along,

\begin{eqnarray}
\label{CURV1}
\epsilon_{ijk}\epsilon^{abc}\hbar{G}{\delta \over {\delta{A}^a_i}}
\Bigl[\hbar{G}\delta^{(3)}(\boldsymbol{x})D^{kj}_{cb}+B^k_{c}(\Psi_{be}B^j_{e})\Bigr]
\bigl\vert\Psi_{GKod}\bigr>\nonumber\\
=\epsilon_{ijk}\epsilon^{abc}\biggl[(\hbar{G}\delta^{(3)}(\boldsymbol{x}))^{2}\epsilon^{kji}_{cba}
+\hbar{G}\delta^{(3)}(\boldsymbol{x}){\partial \over {\partial{A}^a_i}}(B^k_{c}\Psi_{be}B^j_{e})\nonumber\\
+\hbar{G}\delta^{(3)}(\boldsymbol{x})D^{kj}_{cb}(\Psi_{ae}B^i_{e})+B^k_{c}(\Psi_{be}B^j_{e})(\Psi_{af}B^i_{f})\biggr]\bigl\vert\Psi_{GKod}\bigr>
\end{eqnarray}

\noindent
The semiclassical part of (\ref{CURV1}) is given by

\begin{eqnarray}
\label{CUURRVV}
\epsilon_{ijk}\epsilon^{abc}B^k_{c}(\Psi_{be}B^j_e)(\Psi_{af}B^i_f)
=(\hbox{det}B)\epsilon_{fec}\epsilon^{abc}\Psi_{af}\Psi_{be}\nonumber\\
=(\hbox{det}B)\bigl(\delta^a_{f}\delta^b_{e}-\delta^a_{e}\delta^b_{f}\bigr)\Psi_{af}\Psi_{be}
=(\hbox{det}B)Var\Psi.
\end{eqnarray}

The coefficient of the highest degree of singularity $(\hbar{G}\delta^{(3)}(\boldsymbol{x}))^2$ in (\ref{CURV1}) is a nonzero numerical constant equal to 36, as can be seen from the manipulation

\begin{equation}
\label{CURV2}
\epsilon_{ijk}\epsilon^{abc}\epsilon^{kji}_{cba}=(\epsilon_{ijk}\epsilon^{ijk})^2=36.
\end{equation}

\noindent
Therefore, in order to satisfy the quantum Hamiltonian constraint by canonical methods without complications, it is necessary to have a contribution to $\hat{H}$ that cancels this numerical constant.  So the total contribution due to the curvature is given by

\begin{eqnarray}
\hat{H}_{curv}\bigl\vert\Psi_{GKod}\bigr>
=(\hbox{det}B)Var\Psi\nonumber\\
+\hbar{G}\delta^{(3)}(\boldsymbol{x})\Bigl(\epsilon_{ijk}\epsilon^{abc}
{\partial \over {\partial{A}^a_j}}\bigl(B^k_{c}\Psi_{be}B^j_e)
+D^{kj}_{cb}\Psi_{ae}B^i_e\bigr)\Bigr)
+36(\hbar{G}\delta^{(3)}(\boldsymbol{x}))^2
\end{eqnarray}

\par
\indent
Next we move on to the matter contribution.

\subsection{Matter contribution to the expansion of the quantized Hamiltonian constraint}
\par
\medskip
\noindent
The contributions calculated thus far to the quantized Hamiltonian constraint are of the same form regardless of the model, as in the case of the kinematic constraints.  It is the matter contribution that distinguishes one model from another.  The quantized matter contribution to the Hamiltonian constraint for a general matter field will be of the general form

\begin{eqnarray}
\label{MATTERCONTRIBUTION}
G\hat{\Omega}\Psi_{GKod}=\Bigl(G\sum_{n=-\infty}^{\infty}(\hbar{G}\delta^{(3)}(0))^n\Omega_n\Bigr)\Psi_{GKod}
\end{eqnarray}

\noindent
where in (\ref{MATTERCONTRIBUTION}), $\Omega_n$ are the model-specific coefficients for a given degree of singularity and we have made the replacement, in an abuse of notation, of $\delta^{(3)}(\boldsymbol{x})\rightarrow\delta^{(3)}(0)$.  Before attempting to solve the constraints, we must take into account the contributions due to the matter fields for the full theory.\par
\indent  
Let us will illustrate using a Klein--Gordon scalar $\phi$ with conjugate momentum $\pi$.  We will assume that the scalar potential $V(\phi)$ can be included as a contribution to the cosmological 
term $\Lambda$, hence it suffices to consider the kinetic and the spatial gradient terms.  Starting with the classical form of this contribution,

\begin{equation}
\label{KG}
H_{KG}={{\pi^2} \over 2}+{1 \over 2}\partial_{i}\phi\partial_{j}\phi\widetilde{\sigma}^i_{a}\widetilde{\sigma}^j_{a}
\end{equation}

\noindent
we have, upon quantization and making the identification 
$(1/2)\partial_{i}\phi\partial_{j}\phi=T_{ij}$,

\begin{eqnarray}
\label{KGONE}
(\hat{H})_{KG}\Psi_{GKod}[A^a_i,\phi]
=\biggl[-{{\hbar^2} \over 2}{{\delta^2} \over {\delta\phi(x)\delta\phi(x)}}
+\hbar^2{G^2}T_{ij}{\delta \over {\delta{A^a_i(x)}}}
{\delta \over {\delta{A^a_j(x)}}}\biggr]\Psi_{GKod}[A^a_i,\phi]
\end{eqnarray}

\noindent
Continuing on from (\ref{KGONE}) and making use of the CDJ Ansatz,

\begin{eqnarray}
\label{KG1}
(\hat{H})_{KG}\Psi_{GKod}[A^a_i,\phi]
=\biggl[-{{i\hbar} \over 2}{\delta \over {\delta\phi(x)}}\pi(x)
+\hbar{G}T_{ij}{\delta \over {\delta{A^a_i(x)}}}\bigl(\Psi_{ae}(x)B^i_{e}(x)\bigr)\biggr]
\Psi_{GKod}[A^a_i,\phi]\nonumber\\
=\biggl[{{\pi^2} \over 2}+T_{ij}\Psi_{ae}\Psi_{af}B^i_{e}B^j_{f}
+\hbar{G}\delta^{(3)}(0)\biggl(-{i \over {2G}}{{\partial\pi} \over \partial\phi}
+T_{ij}\Bigl(B^i_{e}{\partial \over {\partial{A^j_{a}}}}\Psi_{ae}
+D^{ij}_{ea}\Psi_{ae}\Bigr)\biggr)\biggr]\Psi_{GKod}[A^a_i,\phi]
\end{eqnarray}

\noindent
The term $T_{ij}D^{ij}_{ea}$ in (\ref{KG1}) vanishes due to symmetry of $T_{ij}$ and antisymmetry of $D^{ij}_{ea}$.  One can thus read off from (\ref{KG1}) the contributions to $\Omega_0$ and $\Omega_1$ in (\ref{MATTERCONTRIBUTION}) due to the Klein--Gordon field.  These are given by

\begin{eqnarray}
\label{KG2}
\Omega_0={{\pi^2} \over 2}+T_{ij}\Psi_{ae}\Psi_{af}B^i_{e}B^j_{f}
={{\pi^2} \over 2}+\delta_{ab}\tau_{ef}\Psi_{ae}\Psi_{bf}
;\nonumber\\
\Omega_1=-{i \over {2G}}{{\partial\pi} \over \partial\phi}
+T_{ij}B^i_{e}{\partial \over {\partial{A^j_{a}}}}\Psi_{ae}
=-{i \over {2G}}{{\partial\pi} \over \partial\phi}
+\tau_{ej}{\partial \over {\partial{A^j_{a}}}}\Psi_{ae}
\end{eqnarray}

\noindent
and $\Omega_N=0$ for $N\geq{2}$ and $N<0$, where we have defined $\tau_{ef}=T_{ij}\Psi_{ae}\Psi_{af}B^i_{e}B^j_{f}$ as the undensitized projection of the space-space components of the energy momentum tensor from $\Sigma$ into $SU(2)_{-}$, with $\tau_{ej}=T_{ij}B^i_e$ corresponding to the projection of just one spatial index 
into $SU(2)_{-}$.

\subsection{Putting it all together}
\par
\medskip
\noindent
The full expansion of the quantum Hamiltonian constraint can be written in the form, combining all terms,

\begin{eqnarray}
\label{TOG}
\hat{H}\bigl\vert\Psi_{GKod}\bigr>
=\bigl(H_{\Lambda}+H_{curv}+H_{matter}\bigr)\bigl\vert\Psi_{GKod}\bigr>
=\biggl[\hbox{det}B\bigl(Var\Psi+(\Lambda+GV)\hbox{det}\Psi\bigr)+G\Omega_0\nonumber\\
+\hbar{G}\delta^{(3)}(\boldsymbol{x})\Bigl(\epsilon_{ijk}\epsilon^{abc}
\bigl[D^{kj}_{cb}B^i_e\Psi_{ae}+{\partial \over {\partial{A}^a_i}}\bigl(B^k_{c}B^j_e\Psi_{be}+
{{(\Lambda+GV)} \over 4}B^k_{e}B^j_{f}\Psi_{ce}\Psi_{bf}\bigr)\bigr]+G\Omega_1\Bigr)\nonumber\\
+(\hbar{G}\delta^{(3)}(\boldsymbol{x}))^{2}
\Bigl({{(\Lambda+GV)} \over 6}\epsilon_{ijk}\epsilon^{abc}
{\partial \over {\partial{A^a_i}}}{\partial \over {\partial{A^b_j}}}
(\Psi_{ce}B^k_{e})+36\Bigr)\biggr]\bigl\vert\Psi_{GKod}\bigr>
\end{eqnarray}

\noindent
We see from (\ref{TOG}) that a third-order functional differential condition on the generalized Kodama wavefunction $\Psi_{GKod}$ is equivalent to a second-order partial differential condition on the CDJ matrix elements.  The tradeoff is that whereas the former functional differential condition is linear, the latter partial differential condition is nonlinear.  The expansion (\ref{TOG}) can be written in compact form as

\begin{equation}
\label{HAMMI4}
\hat{H}(\boldsymbol{x})\Psi_{GKod}=\bigl[q_0(\boldsymbol{x})+\hbar{G}\delta^{(3)}(\boldsymbol{x})q_1(\boldsymbol{x})
+(\hbar{G}\delta^{(3)}(\boldsymbol{x}))^{2}q_2(\boldsymbol{x})\bigr]\Psi_{GKod}=0
~\forall{\vec{x}}
\end{equation}

\noindent
In order for (\ref{HAMMI4}) to be satisfied for all $\boldsymbol{x}$ in $\Sigma_T$, which is equivalent to the condition that the quantum Hamiltonian constraint, upon direct promotion from its classical counterpart (which stems from the requirement $\delta{I_{Ash}}/\delta{\underline{N}(x)}=0~\forall{x}\in{M}$), be identically satisfied $\forall\boldsymbol{x}$, we must impose that
$q_{0}(\boldsymbol{x})=0~\forall\boldsymbol{x}$ on $\Sigma_T$.  This is the classical part of the Hamiltonian constraint and also forms the semiclassical part of the SQC.\par
\indent  
Note for $\boldsymbol{x}\neq{0}$ that the quantized Hamiltonian constraint is identically zero due to the delta functions, which have support only at the origin 
$(\boldsymbol{x},t)=(0,t)$ for all times $t$ in $M$.  So there is an automatically manifest semiclassical-quantum correspondence for all points not including the spatial origin.  But we require that the quantized Hamiltonian constraint be satisfied everywhere, including the origin, as a necessary condition for a finite state.  This dictates, and is often put in an abuse of notation, that 

\begin{equation}
\label{HAMMI5}
\hat{H}\bigl\vert\Psi_{GKod}\bigr>=\bigl[q_0+\hbar{G}\delta^{(3)}(0)q_1+(\hbar{G}\delta^{(3)}(0))^{2}q_2\bigr]\bigl\vert\Psi_{GKod}\bigr>=0.
\end{equation}

\indent
The continuity of the SQC imposes conditions on the coefficients of the singular delta functions in (\ref{HAMMI5}), namely that $q_0=q_1=q_2=0$ for all 
$\boldsymbol{x}$ on the hypersurface $\Sigma_t$ for each $t$.  Since the origin of $\Sigma$ can be arbitrarily chosen, then these conditions must be satisfied at all points $\boldsymbol{x}$ in $\Sigma$.  This implies certain functional relationships in the coefficients $q_0$, 
$q_1$ and $q_2$, amongst the fields $A^a_i=A^a_{i}(x)$ and $\phi^{\alpha}=\phi^{\alpha}(x)$ which must be true for all $\boldsymbol{x}$ independently of position 
in $\Sigma$.  The explicit $\boldsymbol{x}$ dependence of the fields themselves can then be suppressed, since $\boldsymbol{x}$ is merely a dummy label.

\section{Constraints corresponding to the finite states of quantum gravity}

The existence of $\Psi_{GKod}$ in the full theory (including minisuperspace as a subset) will depend upon the existence of solutions to the quantum constraints for the CDJ matrix elements for an arbitrary model coupled to gravity with cosmological term.  As long as (\ref{DIFF}),(\ref{QUANT7}) and (\ref{QUIT1}) are satisfied, quantum states obeying (\ref{QUANT3}) and (\ref{ANSATZ2}) are semiclassical in the precise sense that these states will be exponentials (of the Hamiltonian function), and the Hamilton--Jacobi equations will also hold.\par
\indent  
The resulting condition upon the CDJ matrix elements $\Psi_{ae}$ is a system of nine equations in nine unknowns.  This is a total of three equations from the quantized Gauss' law constraints, three equations from the quantized diffeomorphism constraint and three equations from the quantized Hamiltonian constraint.  It will be convenient to think of these nine equations as a map from the nine CDJ matrix elements $\Psi_{ae}$ to the nine equations, thought of as the components of a 
nine-vector $C_{ab}=0$. 

\begin{equation}
\Psi_{ab}(x)\longrightarrow{C}_{ab}[\Psi_{ef}[A^a_i(x)]].
\end{equation}

\noindent
First let us write the system corresponding to the pure Kodama state.

\begin{eqnarray}
\label{SYSKOD}
\epsilon_{aed}\Psi_{ae}=0;\nonumber\\
\Bigl(\delta_{af}{{\partial} \over {\partial{t^g}}}+C_{a}^{fg}\Bigr)\Psi_{fg}=0;\nonumber\\
q_0=\hbox{det}B\bigl(\Lambda\hbox{det}\Psi+Var\Psi\bigr)=0;\nonumber\\
q_1=\epsilon_{ijk}\epsilon^{abc}D^{kj}_{cb}\Psi_{ae}B^i_{e}+
\epsilon_{ijk}\epsilon^{abc}{\partial \over {\partial{A}^a_i}}
\Bigl[B^k_{c}B^j_{e}\Psi_{be}+{\Lambda \over 4}B^k_{e}B^j_{f}\Psi_{ce}\Psi_{bf}\Bigr]=0;\nonumber\\
q_2={\Lambda \over 6}{\partial \over {\partial{A^a_i}}}{\partial \over {\partial{A^b_j}}}
(\epsilon_{ijk}\epsilon^{abc}B^k_{e}\Psi_{ce})+36=0
\end{eqnarray}

\noindent
The system (\ref{SYSKOD}) is a nonlinear system with solution 
$\Psi_{ae}=-6\Lambda^{-1}\delta_{ae}$, which corresponds to the pure Kodama state $\Psi_{Kod}$ in a quantum theory of gravity free of field-theoretical singularities at the level of the state.  Note that all nine degrees of freedom in the CDJ matrix for $\Psi_{Kod}$ are exhausted in order to produce a unique solution.\par
\indent
The basic principle of the nonperturbative quantization of gravity in the general case in the full theory is to introduce a driving force to the right hand side 
of (\ref{SYSKOD}) corresponding to a particular matter model.  The associated criterion of finiteness of the quantum state produces a system which would hopefully converge, in the functional sense, to the CDJ matrix elements $\Psi_{ae}$ for the generalized Kodama state for the model with all of its degrees of freedom similarly exhausted.  In the case of the Klein--Gordon field with self-interaction potential $V=V(\phi(x))$ coupled to gravity the associated system then becomes  

\begin{eqnarray}
\label{SYSGKODD}
\epsilon_{aed}\Psi_{ae}=G\widetilde{\tau}_{0d};\nonumber\\
\Bigl(\delta_{af}{{\partial} \over {\partial{t^g}}}+C_{a}^{fg}\Bigr)\Psi_{fg}=GQ_a=0;\nonumber\\
\hbox{det}B\bigl((\Lambda+GV)\hbox{det}\Psi+Var\Psi\bigr)
+G\bigl({{\pi^2} \over 2}+\delta_{ab}\tau_{ef}\Psi_{ae}\Psi_{bf}\bigr)=0;\nonumber\\
\epsilon_{ijk}\epsilon^{abc}D^{kj}_{cb}\Psi_{ae}B^i_{e}+
\epsilon_{ijk}\epsilon^{abc}{\partial \over {\partial{A}^a_i}}
\Bigl[B^k_{c}B^j_{e}\Psi_{be}+{{(\Lambda+GV)} \over 4}B^k_{e}B^j_{f}\Psi_{ce}\Psi_{bf}\Bigr]\nonumber\\
-{i \over {2}}{{\partial\pi} \over \partial\phi}
+G\tau_{ej}{\partial \over {\partial{A^j_{a}}}}\Psi_{ae}=0;\nonumber\\
{{(\Lambda+GV)} \over 6}{\partial \over {\partial{A^a_i}}}{\partial \over {\partial{A^b_j}}}
(\epsilon_{ijk}\epsilon^{abc}B^k_{e}\Psi_{ce})+36=0
\end{eqnarray}

\noindent
The system (\ref{SYSGKODD}) represents a system of nine equations in nine unknowns, corresponding to the model of the Klein--Gordon field coupled to quantum gravity in the full theory.  Likewise, this system should have a unique solution for its CDJ matrix elements $\Psi_{ae}$ corresponding to its associated generalized Kodama state $\Psi_{GKod}$, when the mixed partials condition is taken into account.\footnote{One could eliminate all components of the  semiclassical matter momentum 
from (\ref{SYSGKODD}) by using the integrated from of the mixed partials condition 
$\pi_{\alpha}(A^a_i,\vec{\phi})=f_{\alpha}(\vec{\phi})
-{i \over G}{\partial \over {\partial\phi^{\alpha}}}
\Bigl(\int_{\Gamma}\delta{X}^{ae}\Bigr)\Psi_{ae}(A^b_j,\vec{\phi})$, $N$ conditions where $N$ is the number of semiclassical matter conjugate momentum components $\pi_{\alpha}$ labeled by $\alpha$, and then look for solutions of the resulting nine by nine system or alternatively, incorporate these $N$ conditions into the constraints themselves dimensionally expanding them to a $9+N$ by $9+N$ system which must be solved.  Note that the generalized Kodama state $\Psi_{GKod}$ then acquires the label $f_{\alpha}$, the functional boundary condition corresponding to the semiclassical matter momentum in the absence of gravity.  It is this label which establishes the link from the quantum gravitational state $\Psi_{GKod}$ to the semiclassical limit below the Planck scale, where the manifestation of quantum gravitational effects may possibly be predicted or ruled out.}  Note that the self interaction potential $V$ can be treated as a contribution to the cosmological constant $\Lambda$.  This procedure should in general be applicable to a wide class of models coupled to quantum gravity.  In the case of the Klein--Gordon field the $SU(2)$ charge 
$Q_a$ is zero.  The shorthand notation for the system (\ref{SYSGKODD}) can be written as a map

\begin{eqnarray}
\label{SYSKOD1}
\Psi_{ab}[A^a_i(x),\phi(x)]\longrightarrow{C}_{ab}[\Psi_{ef}[A^a_i(x),\phi(x)]]\nonumber\\
=O_{ab}^{cd}\Psi_{cd}+\Lambda^{\prime}{I}_{ab}^{cdef}\Psi_{cd}\Psi_{ef}+{\Lambda^{\prime}}^2{E}_{ab}^{cdefgh}\Psi_{cd}\Psi_{ef}\Psi_{gh}=GQ_{ab}
\end{eqnarray}

\noindent
for suitably defined $O_{ab}^{cd}$, ${I}_{ab}^{cdef}$ and ${E}_{ab}^{cdefgh}$, and $Q_{ab}$ where $\Lambda^{\prime}=\Lambda+GV$.\footnote{For a demonstration of the technique for the general solution to the system (\ref{SYSGKODD}), (\ref{SYSKOD1}) the interested reader is directed to \cite{EYOFULL}.}

\section{Theoretical arguments for the existence of generalized Kodama states}

\noindent
The following section is designed to examine the pure and generalized Kodama states from various perspectives relevant to the mathematical proof of their existence.  Section 11.1 shows that the analogue of the topological ${F}\wedge{F}$ Chern--Simons term for gravity when quantized on an equal footing in the presence of matter fields is simply another representation of $\Psi_{GKod}$ upon solution to the quantum constraints.  We derive and contrast this to work of previous authors, which view the CDJ matrix as a Lagrange multiplier necessary to enforce metricity as opposed to our use here.  The presence of matter fields coupled to gravity does not alter the holographic nature of the state.\footnote{This refers to velocity independence as well as independence of the bulk configuration of the fields, a characteristic shared by topological field theories.}.  They simply, in a sense, `rotate' the holographic ground state $\Psi_{Kod}$ into a new holographic ground state $\Psi_{GKod}$.  In section 11.2 we show, using geometric quantization, that the uniqueness of the solution $C_{ab}=0$ to the quantum constraints guarantees the existence 
of $\Psi_{Kod}$ and $\Psi_{GKod}$ via the cohomology of the fibre bundle with base space consisting of the configuration space variables.  The significance of the geometric quantization procedure outlined in section 11.2 is that it implies the equivalence of the reduced phase space and the Dirac canonical quantization procedures.

\subsection{Instanton representation of the generalized Kodama state}

\noindent
We will show in this section an important representation of the generalized Kodama state $\Psi_{GKod}$ in analogy to the instanton ${F}\wedge{F}$ term for the pure Kodama state $\Psi_{Kod}$ generalized to the case when matter fields are present.  We must first show, in analogy to the steps leading from (\ref{PAP6}) to (\ref{PAPTOOT1}), that the matter-coupled case does in fact correspond to a wavefunction defined only on the 3-dimensional boundary $\Sigma_T$ at time $T$, and is in fact independent of the gravitational and matter fields in the bulk $M$ of spacetime.  Complementarily to the matter-free case previously discussed, we shall proceed in a more direct manner.  We will start with the assumption that the wavefunction is a boundary term defined on the final spatial hypersurface $\Sigma_T$ at time $T$, and then explicitly show that this boundary term is the same as its four-dimensional version, the precise form of which we will deduce.  The ability to do so without contradiction should suffice to prove our assertion.\par
\indent
Assume that there exists a functional $\Psi_{GKod}(T)=\Psi_{GKod}[A(T),\vec{\phi}(T)]=e^{I(T)}$ of two dynamical variables $(A^a_i(\boldsymbol{x},T),\phi^{\alpha}(\boldsymbol{x},T))$ as determined by the solution to the constraints (\ref{SYSKOD1}) on the final spatial hypersurface $\Sigma_T$.  

\begin{eqnarray}
\label{WAVY}
\Psi(T)=e^{I(T)}=e^{I(T)-I(t_0)}e^{I(t_0)}=e^{\int^T_{t_0}dt(dI/dt)}\Psi(t_0)
\end{eqnarray}

\noindent  
$\Psi(t_0)=\Psi_{GKod}[A(t_0),\vec{\phi}(t_0)]$ is the state as determined by the solution to the constraints (\ref{SYSKOD1}) on the initial spatial 
hypersurface $\Sigma_0$.\footnote{This is the case because the form of the constraints (\ref{SYSGKODD}), as well as the mixed partials condition, is preserved on each hypersurface $\Sigma_t$.}  On the right hand side in the last line of (\ref{WAVY}), the functional $I(T)$ is extended from $\Sigma_T$ into the interior of $M$ to produce $I(t)$ for $t_0\leq{t}\leq{T}$.\par
\indent  
Note that the right hand side of (\ref{WAVY}) is independent of this detailed time dependence, since the left hand side is independent of any such dependence by the starting assumption.  Expansion of the time derivative in (\ref{WAVY}) while making use of the field-theoretical time variation of $I$ leads to\footnote{Note that this is a total time derivative and hence is still independent of the bulk configuration of the fields, even though the functional $I$ has been extended into the interior of the spacetime region $M$.}

\begin{eqnarray}
\label{WAVY1}
\Psi(T)=
\hbox{exp}\Bigl[\int^T_{t_0}{dt}
\int_{\Sigma}d^3{x}\Bigl({{\delta{I}} \over {\delta{A}^a_i}}\dot{A}^a_i
+{{\delta{I}} \over {\delta\phi}}\dot{\phi}\Bigr)\Bigr]\Psi(t_0)
\end{eqnarray}

\noindent
We now specify the functional form of the state $\Psi(T)$ on the left hand side of (\ref{WAVY1}) by choosing it to correspond to the solution to the 
constraints (\ref{SYSGKODD}) on the final hypersurface $\Sigma_T$.  Since the functional form of the constraints is the same for each $0\leq{t}\leq{T}$, we can extend the functional form of $I(T)=I[A(T),\vec{\phi}(T)]$ on the spacelike boundary $\partial{M}=\Sigma_T$ to the same functional form $I(t)=I[A(t),\vec{\phi}(t)]$ on the spacelike hypersurface $\Sigma_{t}$ for any $t_0\leq{t}\leq{T}$.  That the constraints have the same form independently of position $x$ in $M$ is a result of the classical equations of motion for the Lagrange multipliers which implement the constraints

\begin{eqnarray}
\label{WAVY2}
{{\delta{S}} \over {\delta\underline{N}(x)}}={{\delta{S}} \over {\delta{N^i}(x)}}={{\delta{S}} \over {\delta\theta^a(x)}}=0~~\forall{x}~\hbox{in}~M.
\end{eqnarray}

\noindent
Assuming that the constraints when implemented on the quantum level allow a for a unique solution of (\ref{SYSGKODD}) corresponding to the CDJ matrix 
elements $\Psi_{ae}$, we identify the quantity $I$ in (\ref{WAVY1}) with the Hamilton-Jacobi functional for a WKB state.  Since the semiclassical-quantum correspondence holds by construction, then the WKB state is as well the quantum state.  We can then make the replacements

\begin{eqnarray}
\label{WAVY3}
{{\delta{I}} \over {\delta{A}^a_i(x)}}=(\hbar{G})^{-1}\Psi_{ae}[A^a_i(x),\phi^{\alpha}(x)]B^i_{e}[A(x)]
\end{eqnarray}

\noindent
which merely renames the semiclassical gravitational conjugate momentum based upon the solution to the constraints, and

\begin{eqnarray}
\label{WAVY4}
{{\delta{I}} \over {\delta\phi^{\alpha}(x)}}
={i \over \hbar}\Bigl(-i\hbar{{\delta{I}} \over {\delta\phi^{\alpha}(x)}}\Bigr)
={i \over \hbar}\pi_{\alpha}[A^a_i(x),\phi^{\alpha}(x)]
\end{eqnarray}

\noindent
which renames the semiclassical matter conjugate momentum.  Note that (\ref{WAVY3}) and (\ref{WAVY4}) suffice to imply the mixed partials condition which in turn stems from the requirement that $I(T)$ be the integral of a total time derivative, since by evaluation of cross derivatives we have that

\begin{eqnarray}
\label{WWAA}
{{\delta^2I} \over {\delta{A}^a_i\delta\phi^{\alpha}}}
={{\delta^2I} \over {\delta\phi^{\alpha}\delta{A}^a_i}}
={i \over \hbar}{{\delta\pi_{\alpha}} \over {\delta{A}^a_i}}
=(\hbar{G})^{-1}B^i_e{{\delta\Psi_{ae}} \over {\delta\phi^{\alpha}}}.
\end{eqnarray}

\noindent
The mixed partials condition is necessary in order to eliminate the semiclassical matter conjugate 
momentum $\pi_{\alpha}$ from the constraints (\ref{SYSGKODD}), so that the wavefunctional 
$\Psi_{GKod}=\Psi_{GKod}[A^a_i,\phi^{\alpha}]$ can be expressed completely in terms of the configuration variables $(A^a_i,\phi^{\alpha})$.\footnote{As indicated under footnote 16 in section 6, one may view the mixed partials condition as an extension of the quantum constraints to be solved in conjunction with (\ref{SYSGKODD}) as a $9+N$ by $9+N$ system, where $N$ is the number of mixed partials conditions in integrated form, $N$ of them corresponding to the $N$ components of the matter fields, labeled by the index $\alpha$.}
Also let us note for completeness that the mixed partials condition, (\ref{WAVY4}) and (\ref{WWAA}) stem directly from the exhaustive application of the quantization relations defining the theory.  Equations (\ref{KGCCR}) and (\ref{KGCCCR1}) in conjunction with the CDJ Ansatz are necessary in order to obtain the quantized version of the constraints with associated degrees of singularity, which place conditions on the CDJ matrix (\ref{SYSKOD1}) which are unique to the Hamiltonian of general relativity.  Equation (\ref{STATUS}) guarantees that the wavefunctional has the proper polarization to enable these conditions to the determined, and is the correct boundary term corresponding to these conditions.  Equations (\ref{KGCCR1}) and (\ref{KGCCR2}) simply follow as a consistency condition that follows from the existence of the aforementioned boundary term and its polarization.  It then follows from all these observations that the existence of the generalized Kodama 
state $\Psi_{GKod}$ is a natural consequence of the exhaustive application of the canonical quantization relations of quantum field theory to gravity in Ashtekar variables in the Schr\"odinger representation, which follows from the principle of the SQC.\par  
\indent
Proceeding along by substitution of (\ref{WAVY3}) and (\ref{WAVY4}) into (\ref{WAVY1}) leads to

\begin{eqnarray}
\label{WAVY5}
\Psi(T)=\Psi_{GKod}(T)=\nonumber\\
\hbox{exp}\Bigl[\int^T_{t_0}{dt}
\int_{\Sigma}d^3{x}\Bigl((\hbar{G})^{-1}\Psi_{ae}B^i_e\dot{A}^a_i
+{i \over \hbar}\pi_{\alpha}\dot{\phi}^{\alpha}\Bigr)\Bigr]\biggl\vert_{C_{ab}=0;\vec\pi
=\vec{\pi}_f(\phi)}\Psi(t_0)
\end{eqnarray}

\noindent
where the notation $\vec{\pi}_f(\phi)$ signifies that the mixed partials condition in integrated form has already been incorporated into the solution to the constraints.\par
\indent  
We now derive the instanton representation of $\Psi_{GKod}$, which highlights its relation to topological field theory.  The first thing to note is that the argument of the exponential in (\ref{WAVY5}) is nothing other than starting action (\ref{LAGRAN1}) evaluated on the reduced phase space upon solution to the constraints as is as well the integral of a total time derivative.  Writing the argument as a spacetime integral, we have

\begin{equation}
\label{KOMM}
I=\int_{M}\Bigl((\hbar{G})^{-1}\Psi_{ae}B^i_{a}\dot{A}^e_{i}
+{i \over \hbar}\pi_{\alpha}\dot{\phi}^{\alpha}\Bigr).
\end{equation}

\noindent
To see more clearly the relation between the two terms from a different perspective, it helps to express (\ref{KOMM}) in covariant form.  We have, for the gravitational part,

\begin{eqnarray}
\label{COUGHH}
\Psi_{ae}B^i_{a}\dot{A}^e_{i}=\Psi_{ae}B^i_a\bigl(F^e_{0i}+D_{i}A^e_{0}\bigr)=\epsilon^{ijk}\Psi_{ae}F^a_{jk}F^e_{0i}+\widetilde{\sigma}^i_{e}D_{i}A^e_{0}
\end{eqnarray}

\noindent
where we have made the identification in analogy to $SU(2)$ Yang--Mills theory, that

\begin{equation}
F^a_{\mu\nu}=\partial_{\mu}A^a_{\nu}-\partial_{\nu}A^a_{\mu}+f^{abc}A^b_{\mu}A^c_{\nu}.
\end{equation}

\noindent
To show the steps for the zeroth component,

\begin{equation}
F^e_{0i}=\partial_{0}A^e_{i}-\partial_{i}A^e_{0}+f^{efg}A^f_{0}A^g_{i}
\longrightarrow\partial_{0}A^e_{i}=F^e_{0i}+\partial_{i}A^e_{0}+f^{efg}A^f_{i}A^g_{0}=F^e_{0i}+D_{i}A^e_{0}.
\end{equation}

\noindent
Performing an analogous operation for the matter fields,

\begin{equation}
\label{KOMM1}
\pi_{\alpha}\dot{\phi}^{\alpha}=\pi_{\alpha}D_{0}\phi^{\alpha}-A^a_{0}Q_{a}
\end{equation}

\noindent
In (\ref{KOMM1}) we used the definition of the time component of the covariant derivative

\begin{equation}
(D_{\mu}\phi)^{\alpha}=\partial_{\mu}\phi^{\alpha}+A_{\mu}^{a}(T_a)^{\alpha}_{\beta}\phi^{\beta}
\end{equation}

\noindent
with the $SU(2)_{-}$ charge $Q_a$, given by

\begin{equation}
Q_{a}=\pi_{\alpha}(T_a)^{\alpha}_{\beta}\phi^{\beta}.
\end{equation}

\noindent
Let us first rewrite the last term on the last line of (\ref{COUGHH})

\begin{eqnarray}
\label{COFFFE}
\widetilde{\sigma}^i_{a}D_{i}A^a_{0}
=\widetilde{\sigma}^i_{a}\partial_{i}A^a_{0}+\widetilde{\sigma}^i_{a}f^{abc}A^b_{i}A^c_{0}
=\partial_{i}(\widetilde{\sigma}^i_{a}A^a_0)-A^a_{0}\partial_{i}\widetilde{\sigma}^i_{a}+\widetilde{\sigma}^i_{a}f^{abc}A^b_{i}A^c_{0}
\end{eqnarray}

\noindent
The first term on the right hand side of (\ref{COFFFE}) corresponds to a gauge transformation on the two dimensional boundary $\partial\Sigma$ of three-space $\Sigma$, and we are
left with

\begin{equation}
\partial_{i}(\widetilde{\sigma}^i_{a}A^a_0)-A^a_{0}\bigl(\partial_{i}\widetilde{\sigma}^i_{a}+f^{abc}A^b_{i}\widetilde{\sigma}^i_{c}\bigr)
=\partial_{i}(\widetilde{\sigma}^i_{a}A^a_0)-A^a_{0}D_{i}\widetilde{\sigma}^i_{a}.
\end{equation}

\noindent
Having expressed both the gravitational and the matter contributions in covariant notation, making the extension 
$\epsilon_{ijk}=\epsilon_{0ijk}\rightarrow\epsilon_{\mu\nu\rho\sigma}$ we can upon substitution of (\ref{COUGHH}) rewrite (\ref{KOMM}) in covariant notation as

\begin{eqnarray}
\label{KOMMA}
I=\int^T_{t_0}dt\int_{\Sigma}d^3{x}\Bigl((\hbar{G})^{-1}\Psi_{ae}B^i_{a}\dot{A}^e_{i}
+{i \over \hbar}\pi_{\alpha}\dot{\phi}^{\alpha}\Bigr)\nonumber\\
=\int_{M}\Bigl((\hbar{G})^{-1}\Psi_{ae}{F^a}\wedge{F^b}
+{i \over \hbar}\pi^{\alpha}D_{0}\phi^{\alpha}
+{i \over \hbar}A^a_{0}\bigl(D_{i}\widetilde{\sigma}^i_{a}+Q_a\bigr)\nonumber\\
=\int_{M}\Bigl((\hbar{G})^{-1}\Psi_{ae}{F^a}\wedge{F^b}+{i \over \hbar}\pi^{\alpha}D_{0}\phi^{\alpha}\Bigr)
\end{eqnarray}

\noindent
on account of Gauss' law.  Equation (\ref{KOMMA}) signifies an interaction between the gravitational and the matter field based on $SU(2)_{-}$ gauge invariance.  Without knowledge of the CDJ matrix solution, there would be no physical input at this stage of simplicity different to that from $SU(2)$ Yang--Mills theory coupled to matter.  But there is input from gravity, since the gravitational phase space is smaller than its Yang--Mills counterpart owing to the remaining constraints which determine the CDJ matrix.  Note that in the absence of matter, we have $\phi^{\alpha}=\pi_{\alpha}=0$ and the interactions are no longer present.  Then (\ref{KOMMA}) would reduce to the second Chern class $\int\hbox{tr}{F}\wedge{F}$, a topological invariant 
of $M$, and ultimately to the Chern--Simons action $I_{CS}$ on the spatial 
boundary $\Sigma$ in direct analogy to \cite{TQFT4}.  In the more general case the CDJ matrix acts as a kind of matter-induced metric on $SU(2)_{-}$.  Note that since (\ref{KOMMA}) arose from the same starting functional (\ref{KOMM}) which was chosen to be velocity independent and independent of histories within $M$, it follows that the instanton-like term is as well another representation of the generalized Kodama state, which is as well velocity and history independent and furthermore constitutes the anologue of a topological invariant in the presence of matter fields coupled to gravity

\begin{eqnarray}
\label{INSTANTON}
\Psi_{GKod}(T)=e^{\int_{M}\bigl((\hbar{G})^{-1}\Psi_{ab}{F^a}\wedge{F^b}
+{i \over \hbar}\vec{\pi}\cdot{D_0}\vec{\phi}\bigr)}\Psi_{GKod}(t_0).
\end{eqnarray}

\noindent
Note that one can expand the generalized Kodama instanton representation about its pure Kodama counterpart via the change in 
variables $\Psi_{ab}=-\bigl(6\Lambda^{-1}\delta_{ab}+\epsilon_{ab}\bigr)$, where $\epsilon_{ae}$ parametrizes the departure of $\Psi_{GKod}$ from $\Psi_{Kod}$, to obtain

\begin{eqnarray}
\label{INSTANTON1}
\Psi_{GKod}[A(T),\vec\phi(T)]=e^{6(\hbar{G}\Lambda)^{-1}\int_{M}\hbox{tr}{F}\wedge{F}}
e^{\int_{M}\bigl(-(\hbar{G})^{-1}\epsilon_{ab}{F^a}\wedge{F^b}+{i \over \hbar}\vec{\pi}\cdot{D_0}\vec{\phi}\bigr)}\nonumber\\
=\Psi_{Kod}[A(t_0)]e^{\int_{M}\bigl(-(\hbar{G}\Lambda)^{-1}\epsilon_{ab}{F^a}\wedge{F^b}+{i \over \hbar}\vec{\pi}\cdot{D_0}\vec{\phi}\bigr)}.
\end{eqnarray}

\indent
To go one step further, the gravitational sector of the generalized Kodama state can be taken `off-shell' by the introduction of a self-dual 
two-form $\Sigma^a=\Sigma^a_{\mu\nu}dx^{\mu}\wedge{dx^{\nu}}$.  The `phase' of this sector then becomes

\begin{eqnarray}
\label{SELFDUAL}
-{1 \over 4}\int_M\Psi_{ae}{F^a}\wedge{F^b}
=\int_M\bigl({\Sigma^a}\wedge{F^a}+(\Psi^{-1})_{ab}
{\Sigma^a}\wedge{\Sigma^b}\bigr).
\end{eqnarray}

\noindent
Comparison of (\ref{SELFDUAL}) with \cite{DUALITY1},\cite{DUALITY2} shows a distinct difference in that when matter is quantized with gravity on the same footing, the degrees of freedom in the CDJ matrix $\Psi_{ae}$ are exhausted.  The quantity $(\Psi^{-1})_{ae}$ in (\ref{SELFDUAL}) would have the interpretation in \cite{DUALITY1} as the inverse of the self-dual part of the Weyl tensor in the $SU(2)_{-}$ spinor representation $\Psi_{(ABCD)}$, which serves as a Lagrange-multiplier designed to impose compatability of the instanton-like action with a metric via the identification (in the language of the present paper) ${\Sigma^a}\wedge{\Sigma^b}
={1 \over 3}\delta^{ab}\hbox{tr}{\Sigma}\wedge{\Sigma}$.  However, 
$\Psi_{ae}$ in 
(\ref{SELFDUAL}) has been used to solve the quantum constraints necessary to 
determine $\Psi_{GKod}$ and not as a Lagrange mutiplier, which might in turn be interpreted as a quantum effect on the Weyl curvature tensor due to the presence of the matter fields which cannot be deduced based solely on classical general relativity.  The equation of motion for $\Sigma^a$ enforces the generalized self-duality condition as in \cite{DUALITY1},\cite{DUALITY2}, however the equation of motion for the connection

\begin{eqnarray}
\label{SELFDUAL1}
D\Sigma_a+(\Psi^{-1})_{ce}\Bigl({{\delta\Psi_{ef}} \over {\delta{A^a}}}\Bigr)
(\Psi^{-1})_{fb}{\Sigma^c}\wedge{\Sigma^b}=0
\end{eqnarray}

\noindent
signifies (unlike \cite{DUALITY1},\cite{DUALITY2}) that the Ashtekar connection $A^a_i$ in the presence of matter fields quantized on the same footing with gravity is in general not torsion free.\footnote{This does not in general preclude the existence of a metric with which $A^a_i$ is compatible.}

\subsection{Existence of $\Psi_{GKod}$ from the perspective of geometric quantization vis-a-vis the SQC}

\noindent
The basic notion for the time-parametrization invariance of the generalized Kodama states can also be seen from the application of geometric quantization to constrained systems.  We will demonstrate the relationship to the semiclassical-quantum correspondence and how this can lead uniquely to such independence.  In a more mathematically precise sense, the relationship between geometric quantization and the SQC is none other than the equivalence of the reduced phase space and Dirac quantization of a quantum system,\footnote{Pointed out in \cite{TQFT4} as a requirement for the quantization of a topological theory on a spacetime manifold $M$ to lead to a unique quantum state living on the boundary $\partial{M}$} which exists only for special states including $\Psi_{Kod}$ and 
$\Psi_{GKod}$.\par
\indent
Let us start with the the determination of the pure Kodama 
state $\Psi_{Kod}$.  One starts first with a symplectic 
structure $\boldsymbol{\Sigma}=(\boldsymbol{\omega},\boldsymbol{M})$ with symplectic (functional) 2-form $\boldsymbol{\Omega}$ on the classical phase space 
$\boldsymbol{M}\equiv(\widetilde{\sigma}^i_a(x),A^a_i(x))$ on the configuration space of connections $\Gamma_A$ living at the point $x\in{M}$.  The symplectic 2-form on $\boldsymbol{\Sigma}$ associated to a spatial hypersurface $\Sigma_t$ labeled by time $t$ is given by

\begin{eqnarray}
\label{SIMPLE}
\boldsymbol{\omega}_{\boldsymbol{M}}=(\hbar{G})^{-1}\int_{\Sigma}d^3\boldsymbol{x}
~{\delta\widetilde{\sigma}^i_a(x)}\boldsymbol{\wedge}{\delta{A}^a_i(x)},
\end{eqnarray}

\noindent
where the wedge product is taken with respect to the cotangent bundle of the fields at each point $x$.  At the level of (\ref{SIMPLE}) the phase space per point is $18$ dimensional.  Hence the associated volume form corresponding to (\ref{SIMPLE})

\begin{eqnarray}
\label{VOLUME}
d\mu(A)=w[A]\prod_{a,i}{\delta\widetilde{\sigma}^i_a(x)}\boldsymbol{\wedge}{\delta{A}^a_i(x)}
\end{eqnarray}

\noindent
is nonzero.  Note that (\ref{VOLUME}) signifies the 18-dimensional phase space volume per point $x$, and $w[A]$ is the weighting factor used 
in (\ref{STATTTES}).  The volume form in (\ref{VOLUME}) is nonzero, which in the language of geometric quantization corresponds to a nondegenerate two form $\boldsymbol{\omega}_{M}$.\par
\indent
At the classical level there are seven constraints, which upon solution leave a two-parameter ambiguity in the CDJ matrix $\Psi_{ae}$.  This corresponds to a two-parameter ambiguity in the conjugate 
momentum $\widetilde{\sigma}^i_a=\Psi_{ae}B^i_e$ at the level of the reduced phase space $\boldsymbol{m}$.  Denote the undetermined momenta, without loss of generality, by 
$\widetilde{\sigma}^1_1$ and $\widetilde{\sigma}^2_2$.  The volume form on the original (18 dimensional) phase space $\boldsymbol{M}$ is zero, which means that the corresponding two form 
$\boldsymbol{\omega}_{\boldsymbol{M}}$, though not zero, is degenerate on the space $\boldsymbol{M}$.  However, it is nondegenerate on the reduced (four dimensional) phase space 
${\boldsymbol{m}}$, with a simplectic 2-form given by

\begin{eqnarray}
\label{SIMPLE1}
\boldsymbol{\omega}_{\boldsymbol{m}}=(\hbar{G})^{-1}\int_{\Sigma}d^3\boldsymbol{x}
\Bigl[{\delta\widetilde{\sigma}^1_1(x)}\boldsymbol{\wedge}{\delta{A}^1_1(x)}
+{\delta\widetilde{\sigma}^2_2(x)}\boldsymbol{\wedge}{\delta{A}^2_2(x)}\Bigr]
\neq{0}.
\end{eqnarray}

\noindent
The volume form on this reduced (eleven dimensional) phase space per point is non zero, and given by

\begin{eqnarray}
\label{VOLUME}
d\mu_{red}(A)=w[A]{\delta\widetilde{\sigma}^1_1(x)}\boldsymbol{\wedge}{\delta{A}^1_1(x)}\wedge{\delta\widetilde{\sigma}^2_2(x)}\boldsymbol{\wedge}{\delta{A}^2_2(x)}\prod_{b,j\neq{1,2}}\delta{A}^b_j
\end{eqnarray}

\noindent
This ambiguity in the state of the system presents an obstruction to progress from the prequantization to the quantization stage of geometric quantization in that some momenta would still remain in the wavefunction, assuming that it can be constructed.  Labeling this by $(\Psi_{\boldsymbol{m}})_{Wkb}$ to signify the classical level of solution, one has

\begin{eqnarray}
\label{SIMPLE2}
(\Psi_{\boldsymbol{m}})_{Wkb}
=\Psi_{\boldsymbol{m}}\bigl[\widetilde{\sigma}^1_1(x),
\widetilde{\sigma}^2_2(x),A^a_i(x)\bigr].
\end{eqnarray}

\noindent
Equation (\ref{SIMPLE2}) as it stands is unsuitable for a wavefunction of the universe, since one requires a polarization for which the wavefunction depends completely on `coordinate' variables $A^a_i$.  Indeed, the requirement of integrability is not met at the naive classical level and it is not in general clear the manner in which to construct a state on the 
boundary $\partial{M}$ of spacetime.\par
\indent
However, we will show that the full solution to (\ref{SYSKOD}) at the quantum level removes this ambiguity, enabling the corresponding quantum state to be explicitly constructed.  Our argument proceeds as follows.  Upon full solution to the quantum constraints, all momenta, including 
$\widetilde{\sigma}^1_1$ and $\widetilde{\sigma}^2_2$ are completely determined by the coordinate variables $A^a_i$.  The interpretation is that the symplectic two-form $\boldsymbol{\omega}_{\boldsymbol{m}}$ in 
(\ref{SIMPLE1}) collapses to zero.\footnote{This is the field-theoretical analogy to finite dimensional systems.  Take the harmonic oscillator with $\omega={dp}\wedge{dq}$.  Upon solution to the constraint of constant energy $2E=p^2+q^2$ the symplectic two-form reduces to 
$\omega=(2E-q^2)^{-1/2}q{dq}\wedge{dq}=0$.}  This means that the reduced phase space has decreased from $2\otimes\infty$ degrees of freedom per point to $0\otimes\infty$ degrees of freedom per point.\par
\indent
A way to see this is to compute the full original 2-form (\ref{SIMPLE}) subject to the solution of the quantum constraints (all nine $C_{ab}=0$ equations).  In this case $\boldsymbol{\omega}_{\boldsymbol{M}}$ on the full phase space collapses into $\boldsymbol{\omega}_{\Gamma_A}$
 on the configuration space.  Suppressing $x$ dependence, we have

\begin{eqnarray}
\label{SIMPLER}
\boldsymbol{\omega}_{\Gamma}=
\boldsymbol{\omega}_{\boldsymbol{M}}\Bigl\vert_{C_{ab}=0}
=(\hbar{G})^{-1}\int_{\Sigma}d^3\boldsymbol{x}
~{\delta\widetilde{\sigma}^i_a(x)}\boldsymbol{\wedge}{\delta{A}^a_i(x)}
\biggl\vert_{C_{ab}=0}\nonumber\\
\end{eqnarray}

\noindent
Expanding this out, noting that 
$\widetilde{\sigma}^i_a(x)=\widetilde{\sigma}^i_a[A^b_j(x)]$ gives

\begin{eqnarray}
\label{SIMPLER1}
(\hbar{G})^{-1}\int_{\Sigma}d^3\boldsymbol{x}
~{\delta\widetilde{\sigma}^i_a(x)}\boldsymbol{\wedge}{\delta{A}^a_i(x)}=
(\hbar{G})^{-1}\int_{\Sigma}d^3\boldsymbol{x}
{{\partial\widetilde{\sigma}^i_a} \over {\partial{A}^b_j}}
{\delta{A}^b_j}\wedge{\delta{A}^a_i}\nonumber\\
=\int_{\Sigma}d^3\boldsymbol{x}
\Bigl({{\partial^2{I}_{Kod}} \over {\partial{A}^a_i\partial{A}^b_j}}\Bigr)
{\delta{A}^b_j}\wedge{\delta{A}^a_i}=0.
\end{eqnarray}

\noindent
Equation (\ref{SIMPLER1}) vanishes due to the symmetric indices in the double derivative being contracted with antisymmetric indices in the wedge product.\par
\indent
To show explicitly the manner in which $\Psi_{Kod}$ arises, we argue in analogy to the cohomology of manifolds.  A closed one-form $\theta$ on a manifold $M$, such that $d\theta=0$ is locally exact by the Poincare Lemma \cite{NAKAMURA}.  Hence $\theta=d\lambda$ for some zero form $\lambda$.  So the manifold $M$ can then be divided into cohomology classes $H^1(M)$ of closed-modulo exact forms.  We now extend this to field theory in the following manner.  As noted in \cite{ASH1} the symplectic 
two-form $\boldsymbol{\omega}_{\boldsymbol{M}}$ can be seen as the abelian curvature of a connection one-form, namely the symplectic (canonical) one-form (the analogy to $\theta=pdq$ for the harmonic oscillator) for 
a $U(1)$ line bundle, given by

\begin{eqnarray}
\label{SIMPLE3}
\boldsymbol{\theta}_{\boldsymbol{M}}
=(\hbar{G})^{-1}
\int_{\Sigma}d^3{x}~\widetilde{\sigma}^i_a(x)\delta{A}^a_i(x).
\end{eqnarray}

\noindent
Equation (\ref{SIMPLE3}) is a one-form from the perspective of the functional space $\Gamma_A$ of fields $A^a_i$, but is actually a four-form on spacetime $M$ integrated over all of three-space $\Sigma$.  It also corresponds to the sympletic form on the original classical phase space, prior to implementation of the constraints.\par
\indent
But we have shown that $\boldsymbol{\omega}_{\boldsymbol{M}}$ vanishes, which corresponds to the unique solution to the quantum constraints being selected from the two-parameter family of semiclassical solutions, or the self-duality condition 
$\widetilde{\sigma}^i_a=-6\Lambda^{-1}B^i_a$ at each point $x$.  The canonical one-form on the reduced phase space then becomes  

\begin{eqnarray}
\label{SIMPLE4}
(\boldsymbol{\theta}_{\boldsymbol{m}})_{Kod}
=-6(\hbar{G}\Lambda)^{-1}\int_{\Sigma}d^3{x}~B^i_a(x)\delta{A}^a_i(x).
\end{eqnarray}

\noindent
This enables the integrability condition to be satisfied, in analogy to the cohomology of finite dimensional spaces.  Hence, since 
$\boldsymbol{\theta}_{\boldsymbol{m}}$ is closed
($\boldsymbol{\omega}_{\boldsymbol{m}}=\delta\boldsymbol{\theta}_{\boldsymbol{m}}=0$), then it must be `locally' exact.\footnote{Locally in the functional sense on the space of functions per point $x$.}  Hence it must be the case that $\boldsymbol{\theta}_{\boldsymbol{m}}
=\delta\boldsymbol{\lambda}$ for some `functional' zero-form 
$\boldsymbol{\lambda}$.\footnote{A zero-form in the sense of a three-form integrated over all 3-space $\Sigma$.}  We argue that the determination of this form $\boldsymbol{\lambda}$ amounts to finding the wavefunction of the universe, and is the direct analogue of 
finding $\boldsymbol{\lambda}\in\wedge^0(M)$ in the finite dimensional case.  The latter can be found directly via the 
integration $\boldsymbol{\lambda}=\int\boldsymbol{\theta}$.  The infinite dimensional analogue is to integrate (\ref{SIMPLE4}) over the functional space of fields, which requires that this functional integration commutes with the spatial integration $d^3x$ over $\Sigma$.  Hence,

\begin{eqnarray}
\label{SIMPLE5}
\boldsymbol{\lambda}_{Kod}=\int_{\Gamma_A}
(\boldsymbol{\theta}_{\boldsymbol{m}})_{Kod}
=-6(\hbar{G}\Lambda)^{-1}
\int_{\Gamma}\int_{\Sigma}d^3{x}~B^i_a(x)\delta{A}^a_i(x)\nonumber\\
=-6(\hbar{G}\Lambda)^{-1}\int_{\Sigma}d^3{x}\int_{\Gamma}B^i_a(x)\delta{A}^a_i(x)\nonumber\\
=-6(\hbar{G}\Lambda)^{-1}\int_{\Sigma}\Bigl({A}\wedge{dA}
+{2 \over 3}{A}\wedge{A}\wedge{A}\Bigr)
=-6(\hbar{G}\Lambda)^{-1}I_{CS}[A].
\end{eqnarray}

\indent
There are a few points to note concerning (\ref{SIMPLE5}).  (i) First, it corresponds to an indefinite functional integral over $\Gamma_A$.  Limits of functional integration can be associated with an initial time $t_0$ and a final time $T$, determining the `evolution' of the state between these two times irrespective of histories.  (ii) Secondly, since the self-duality condition 
$\widetilde{\sigma}^i_a(x)=-6(\Lambda)^{-1}B^i_a(x)$ arose from the Dirac canonical quantization procedure, the reduced phase space quantization and Dirac quantization procedures are equivalent for the state, which can be written as 
$\Psi_{Kod}=e^{\boldsymbol{\lambda}_{Kod}}$, where 
$\boldsymbol{\lambda}_{Kod}\in\boldsymbol{\wedge}^0(\boldsymbol{M})$ is an element of the functional zero-forms on the configuration space of connections.  The pure Kodama state then has the interpretation of the section of the line bundle for which the configuration 
space $\Gamma_A$ comprises the base space. (iii) This is a direct consequence of the SQC which resulted in the degeneracy of the canonical two-form $\boldsymbol{\omega}$ on the reduced phase space, which is zero-dimensional per point $x$ owing to the uniqueness of the state.\footnote{We will call this the infinite-dimensional functional analogue of the Poincare Lemma, for lack of a more descriptive term.}  (iv) It can be seen the relationship between functional and time integration which is independent of spatial integration.  Also, one is free to select the Lorentzian or the Euclidean form of the state via the replacement 
$\boldsymbol{\lambda}_{Kod}\leftrightarrow{i}\boldsymbol{\lambda}_{Kod}$.\par
\indent
We now argue that the preceding argument extends by direct analogy to the generalized Kodama states $\Psi_{GKod}$.  We will highlight the relevant differences as we proceed.  One starts first with a symplectic 
structure $\boldsymbol{\Sigma}^{\prime}=(\boldsymbol{\omega}^{\prime},\boldsymbol{M}^{\prime})$ with symplectic (functional) 2-form 
$\boldsymbol{\Omega}^{\prime}$ on the classical phase space 
$\boldsymbol{M}^{\prime}
\equiv(\widetilde{\sigma}^i_a(x),A^a_i(x)\phi^{\alpha}(x),\pi_{\alpha}(x))$ on the configuration space of connections and matter fields 
$\Gamma\equiv(\Gamma_A,\Gamma_{\phi}$) living at the point $x\in{M}^{\prime}$.  The symplectic 2-form on $\boldsymbol{\Sigma}^{\prime}$ associated to a spatial hypersurface $\Sigma_t$ labeled by time $t$ is given by

\begin{eqnarray}
\label{COMPLIC}
\boldsymbol{\omega}^{\prime}_{\boldsymbol{M}^{\prime}}=\int_{\Sigma}d^3\boldsymbol{x}
\Bigl[(\hbar{G})^{-1}{\delta\widetilde{\sigma}^i_a(x)}\boldsymbol{\wedge}{\delta{A}^a_i(x)}
+{i \over \hbar}{\delta\pi_{\alpha}(x)}\wedge{\delta\phi^{\alpha}(x)}\Bigr]
\end{eqnarray}

\noindent
where the wedge product is taken with respect to the cotangent space of the fields at each point $x$.  At the level of (\ref{COMPLIC}) the configuration space per point is $18+2N$ dimensional, where $N$ is the total number of fields (components labeled by the index $\alpha$) due to the presence of matter.  At the classical level there are seven constraints, which upon solution leave a two-parameter ambiguity in the CDJ matrix $\Psi_{ae}$.  This corresponds to a two-parameter ambiguity in the conjugate momentum $\widetilde{\sigma}^i_a$ at the level of the reduced phase space $\boldsymbol{m}$ for each value of $f_{\alpha}(\phi^{\beta})$ determined by the mixed partials condition.  In analogy to $\Psi_{Kod}$ this constitues an obstruction to proceeding from the prequantization into the quantization stage, in that the sympletic two form does not in general vanish.\par
\indent
By solution of the system (\ref{SYSGKODD}) one eliminates the two-parameter ambiguity to obtain the CDJ 
matrix $\Psi_{ae}=\Psi_{ae}[A^a_i,f_{\alpha},\phi^{\alpha}]$, labeled 
by $f_{\alpha}$, as a function of the configuration space 
variables $(A^a_i,\phi^{\alpha})$.  Hence one bypasses the semiclassical level to arrive directly at the conditions delinieating the unique quantum state $\Psi_{GKod}$.\par
\indent
In order to be able to construct $\Psi_{GKod}$, we must show that the symplectic two-form $\boldsymbol{\omega}^{\prime}_{\boldsymbol{m}^{\prime}}$ vanishes on the reduced phase space $\boldsymbol{m}^{\prime}$.  It is true that there is an ambiguity due to the choice of $f_{\alpha}$, however this does not affect the determination of the symplectic two-form 
since $f_{\alpha}=f_{\alpha}(\vec\phi)$ is a function purely of the matter fields $\phi^{\alpha}$ and not of any momenta.\footnote{One possible criteria for the elimination of this freedom in $\vec{f}$ is to demand that it correspond to $\delta\Gamma_{eff}[\phi]/\delta\phi^{\alpha}$ for the effective action $\Gamma_{eff}[\vec{\phi}]$ of the theory of matter quantized in the absence of gravity ($\hbox{lim}_{G\rightarrow{0}}$), such as to produce the correct semiclassical limit.  We motivate this concept briefly in \cite{EYOFULL},\cite{EYO1}}  The full solution to the constraints (\ref{SYSGKODD}) should result in the determination

\begin{eqnarray}
\widetilde{\sigma}^i_a
=\widetilde{\sigma}^i_a(A^a_i,f_{\alpha},\phi^{\alpha});~~
\pi_{\alpha}=\pi_{\alpha}(A^a_i,f_{\alpha},\phi^{\alpha}),
\end{eqnarray}

\noindent
or the conjugate momenta, labeled by $f_{\alpha}$, in terms of the configuration space variables.  Taking this dependence into account, suppressing the label $f_{\alpha}$, we compute (\ref{COMPLIC}) on the reduced phase space $\boldsymbol{m}^{\prime}$.\footnote{Since all field-theoretical infinities have been factored out in the solution to the constraints functional derivatives can be thought of as partial derivatives at the same spatial point, resembling minisuperspace but in fact still the full theory \cite{EYOCALCULUS}.  We maintain the functional notation on the one-forms for clarity.}

\begin{eqnarray}
\label{COMPLIC1}
\boldsymbol{\omega}^{\prime}_{\boldsymbol{m}^{\prime}}=\int_{\Sigma}d^3\boldsymbol{x}
\Bigl[(\hbar{G})^{-1}{\delta\widetilde{\sigma}^i_a(x)}\boldsymbol{\wedge}{\delta{A}^a_i(x)}
+{i \over \hbar}{\delta\pi_{\alpha}(x)}\wedge{\delta\phi^{\alpha}(x)}\Bigr]\nonumber\\
=\int_{\Sigma}\biggl[(\hbar{G})^{-1}{{\partial{\sigma}^i_a} \over {\partial{A}^b_j}}
{\delta{A^b_j}}\wedge{\delta{A^a_i}}
+(\hbar{G})^{-1}{{\partial\widetilde{\sigma}^i_a} \over {\partial\phi^{\alpha}}}
{\delta\phi^{\alpha}}\wedge{\delta{A}^a_i}
+{i \over \hbar}{{\partial\pi_{\alpha}} \over {\partial{A}^b_j}}
{\delta{A}^b_j}\wedge{\delta\phi^{\alpha}}
+{i \over \hbar}{{\partial\pi_{\alpha}} \over {\partial\phi^{\beta}}}
{\delta\phi^{\alpha}}\wedge{\delta\phi^{\beta}}\biggr].
\end{eqnarray}

\noindent
The terms of (\ref{COMPLIC1}) quadratic in the fields 
(${\delta{A}}\wedge{\delta{A}}+{\delta\phi}\wedge{\delta\phi}$) vanish for the same reason that ${\delta{A}}\wedge{\delta{A}}$ vanished for the pure Kodama state.  This is the infinite dimensional analogue to 
${dx}\wedge{dx}+{dy}\wedge{dy}=0$ for independent variables $x$ and $y$. Expanding out the quadratic terms in direct analogy to (\ref{SIMPLER1}) leads to

\begin{eqnarray}
\label{SIMPLEST1}
(\hbar{G})^{-1}\int_{\Sigma}d^3\boldsymbol{x}
{{\partial\widetilde{\sigma}^i_a} \over {\partial{A}^b_j}}
{\delta{A}^b_j}\wedge{\delta{A}^a_i}
+{i \over \hbar}\int_{\Sigma}d^3\boldsymbol{x}
{{\partial\pi_{\alpha}} \over {\partial\phi^{\beta}}}
{\delta\phi^{\alpha}}\wedge{\delta\phi^{\beta}}
\nonumber\\
=\int_{\Sigma}d^3\boldsymbol{x}
\Bigl({{\partial^2{I}_{Kod}} \over {\partial{A}^a_i\partial{A}^b_j}}\Bigr)
{\delta{A}^b_j}\wedge{\delta{A}^a_i}+
\Bigl({{\partial^2{I}_{GKod}} \over {\partial\phi^{\alpha}\partial\phi^{\beta}}}\Bigr){\delta\phi^{\alpha}}\wedge{\delta\phi^{\beta}}
=0
\end{eqnarray}

\noindent
due to symmetric indices on the double partial derivatives being contracted with antisymmetric indices on due to the wedge product of the two forms.
 Hence all that remain are the cross terms ${\delta{A}}\wedge{\delta\phi}$.  We argue that these terms vanish due to the Mixed partials consistency condition.  Expanding these terms from (\ref{COMPLIC1}) leads to

\begin{eqnarray}
\label{COMPLIC2}
\int_{\Sigma}d^3x\Bigl(
(\hbar{G})^{-1}{{\partial\widetilde{\sigma}^i_a} \over {\partial\phi^{\alpha}}}
{\delta\phi^{\alpha}}\wedge{\delta{A}^a_i}
+{i \over \hbar}{{\partial\pi_{\alpha}} \over {\partial{A}^a_i}}
{\delta{A}^a_i}\wedge{\delta\phi^{\alpha}}
\Bigr)\nonumber\\
=\int_{\Sigma}d^3x\Bigl[
{{\partial^2{I}_{GKod}} \over {\partial{A}^a_i\partial\phi^{\alpha}}}
-{{\partial^2{I}_{GKod}} \over {\partial{A}^a_i\partial\phi^{\alpha}}}\Bigr]{\delta{A}^a_i}\wedge{\delta\phi^{\alpha}}=0
\end{eqnarray}

\noindent
The end result is that the symplectic two-form
$\boldsymbol{\omega}^{\prime}_{\boldsymbol{m}^{\prime}}=0$.  It then follows that the corresponding canonical one-form is an exact functional differential $\boldsymbol{\theta}^{\prime}_{\boldsymbol{m}^{\prime}}
=\delta\boldsymbol{\lambda}_{GKod}$ for some functional zero-form
 $\boldsymbol{\lambda}_{GKod}=\boldsymbol{\lambda}_{GKod}[A,\vec\phi]$.  The generalized Kodama state is then given by the exponential of this zero-form

\begin{eqnarray}
\label{COMPLIC2}
\Psi_{GKod}
=e^{\int\boldsymbol{\theta}^{\prime}_{\boldsymbol{m}^{\prime}}}
=e^{\boldsymbol{\lambda}_{GKod}}.
\end{eqnarray}

\noindent
Hence, the generalized Kodama state is determined by the integral of the canonical one-form corresponding to the matter-coupled theory.  The arguments of this section support the concept that the generalized Kodama state is indeed the generalization of the pure Kodama state from another perspective.

\section{Discussion and future directions}

\noindent
We have delineated a criterion for finiteness of the states of the quantum theory of gravity in the full theory by the introduction of the semiclassical-quantum correpondence.  This can be thought of as the search for quantum states which due to cancellation of field-theoretical infinities are also $WKB$ states.  The significance of this work is to show that the generalized Kodama states $\Psi_{GKod}$ are a special class of states of the full theory of quantum gravity, the direct analogue of the pure Kodama state $\Psi_{Kod}$ in the presence of matter fields.  The existence of the generalized Kodama states is a direct consequence of the consistent and exhaustive and application of the canonical commutation relations to quantum gravity in Ashtekar variables in the functional Schr\"odinger representation.  
\par
\indent  
We have also shown how $\Psi_{GKod}$ is restricted to any chosen spatial hypersurface $\Sigma_T$ of spacetime in the presence of matter fields, in analogy to a similar holographic effect for the pure Kodama states $\Psi_{Kod}$.  The ability to construct the wavefunction $\Psi_{GKod}$ hinges crucially upon the ability to obtain solutions to the set of nine equations $C_{ab}=0$ for the CDJ matrix elements $\Psi_{ae}$ for the full theory stemming from the quantum constraints.  The integrated form of the mixed partials condition, a consistency condition arising from the quantization procedure, must be solved in conjunction with this system in order to obtain a wavefunction solely of configuration space variables defined on the final spatial hypersurface $\Sigma_T$ of spacetime $M=\Sigma\times{R}$.  The boundary condition $f_{\alpha}=f_{\alpha}(\vec{\phi})$ on this mixed partials condition provides a possible link from quantized gravity to its associated semiclassical limit in the case of the generalized Kodama states $\Psi_{GKod}$.\par
\indent
We have also provided a variety of representations for the generalized Kodama states 
$\Psi_{GKod}$ including arguments from geometric quantization, the natural generalization of the Chern--Simons instanton term to include matter fields, as well as arguments for independence of field bulk configurations.  The principle of the SQC establishes that the reduced phase space, Dirac and geometric quantization procedures unambiguously produce the same quantum state for the generalized Kodama states by construction.  This is significant in that the usual ambiguities inherent in the quantization of a given theory via different quantization schemes are absent in the case of the generalized Kodama states.  It is anticipated that this feature should facilitate examination of the semiclassical limit of the quantum theory for these states.\par
\indent  
A next step is to establish equivalence of the three quantization procedures considered in this paper to the path integral quantization procedure introduced in \cite{EYOPATH}.  In this way we hope to show how several issues in quantum gravity can be resolved using these states.  Also, we hope to illustrate in greater detail the specific algorithm to construct solutions to the constraints by inspection, both for minisuperspace and for the full theory, and from these solutions demonstrate the explicit construction of the generalized Kodama states for a wide class of models.  Additional future directions will include but are not limited to those mentioned in the introduction.

\section{Acknowledgements}

\noindent
I would like to than Ed Anderson for his open-mindedness, guidance, advice and support during various lenghty discussions regarding the initial drafts of my work, as well as advice in various aspects of paper writing, as well as his patience and recommendations in reading over the final version.  I would also like to thank various other members of DAMTP.

\end{document}